# Application of K-integrals to radiative transfer in layered media

**Adrian C Selden*[+]**

## Abstract

Simple yet accurate results for radiative transfer in layered media with discontinuous refractive index are obtained by the method of K-integrals, originally developed for neutron transport analysis. These are certain weighted integrals applied to the angular intensity distribution at the refracting boundaries. The radiative intensity is expressed as the sum of the asymptotic angular intensity distribution valid in the depth of the scattering medium and a transient term valid near the boundary. Integral boundary conditions are obtained from the vanishing of the K-integrals of the boundary transient, yielding simple linear equations for the intensity coefficients (two for a half-space, four for a slab or an interface), enabling the angular emission intensity and the diffuse reflectance (albedo) and transmittance of the scattering layer to be calculated. The K-integral method is orders of magnitude more accurate than diffusion theory and can be applied to scattering media with a wide range of scattering albedoes. For example, near five figure accuracy is obtained for the diffuse reflectance of scattering layers of refractive index n = 1.5 with single scattering albedo $\varpi$ in the range $\varpi \in [0.3, 1]$

*20 Wessex Close, Faringdon, Oxfordshire SN7 7YY, UK

[+]adrian_selden@yahoo.com

## 1.Introduction

Of the numerous methods developed for correcting the error involved in calculating the radiative flux at the boundary of a scattering medium using the diffusion equation most apply standard corrections to the diffusion solution e.g. the extrapolated boundary approximation [1]. A few employ boundary corrections derived from the radiative transfer equation, which require separate computation [2, 3]. When the angular intensity distribution $I(\mu)$ near the boundary is required ($\mu$ is the direction cosine), solutions of the transfer equation valid in the boundary region (spatial transients) are added to the asymptotic intensity distribution valid in the depth of the medium far from the boundary [3, 4]. The combined solution is matched to the boundary conditions, enabling evaluation of the relative amplitudes of the transient and asymptotic terms. In the PN approximation $I(\mu)$ is expressed as a sum of Legendre polynomials $P_n(\mu)$ in the range $n \in |0, N|$. This is inserted in the boundary equation and evaluated for a sequence of cosines $\mu_n$ (collocation points for $n \in |0, N|$), requiring matrix inversion to determine the polynomial and attenuation coefficients [5]. The diffusion equation involves the zero-order (scalar flux $\varphi$) and first-order (vector flux $\mathbf{J}$) moments of $I(\mu)$, derived from the radiative transfer equation; the vector flux $\mathbf{J}$ is related to the scalar flux gradient via Fick's law: $\mathbf{J} = -D\nabla\varphi$, where D is the diffusion coefficient [6].

An alternative approach, known as the K-integral method, employs weighted moments of the radiative intensity $I(\mu)$ combined with integrated boundary equations [4]. In the K-integral method, unlike the diffusion equation (or the more precise Fokker-Planck equation), the derivation is both exact and independent of the boundary conditions. The approximation arises in the choice of boundary transient, successive approximations yielding increasing accuracy. The use of integrated boundary equations greatly simplifies the calculation, requiring just two simultaneous equations for evaluating the relative intensities of the transient and asymptotic terms, the resulting coefficients being averaged over the angular intensity distribution $I(\mu)$. The results can be highly accurate when compared with radiative transfer theory. The principal advantage of the K-integral method lies in its straightforward application to some challenging problems in radiative transfer not readily tackled by diffusion theory, such as anisotropic scattering in lossy media.

The K-integral method is outlined in the following Section (Sec. 2), with reference to its application to one-dimensional semi-infinite (half-space) and finite width (slab) isotropically scattering media with Fresnel reflection at the dielectric boundaries. The results of albedo calculations are discussed in Sec. 3 and compared with accurate radiative transfer data from the literature. The merits of the K-integral approach are summarised in Sec. 4. Anisotropic scattering (and an asymptotic analysis of the isotropic scattering problem) are discussed in the Appendices.

## 2. Theory

### 2.1 K-integrals

A brief outline of the K-integral method is given here, following the treatment in [4]. The radiative transfer equation (RTE) for the radiant intensity I(x, μ) at x with direction cosine μ is

$$\mu\frac{\partial I(x,\mu)}{\partial x}+I(x,\mu)=\varpi\int_{-1}^{1}d\mu' f(\mu'\to\mu)I(x,\mu')+S(x,\mu) \qquad (1)$$

where ϖ is the single scattering albedo, f(μ'→ μ) the azimuthally averaged probability of a photon being scattered from μ' to μ (the phase function) and S(x,μ) the source function. The K-integrals are defined as weighted moments of I(x, μ) [7, 8]

$$K_i(x)=\int_{-1}^{1}d\mu\mu g_i(\mu)I(x,\mu) \qquad (2)$$

with weight-functions $g_i(\mu)$ for i = 1, 2

$$g_1(\mu)=\frac{1}{2}(g(\mu)+g(-\mu)) \qquad (3a)$$

$$g_2(\mu)=\frac{1}{2\lambda}(g(\mu)-g(-\mu)) \qquad (3b)$$

and g(μ) satisfies the characteristic equation of radiative transfer

$$(1-\lambda\mu)g(\mu)=\varpi\int_{-1}^{1}d\mu' f(\mu\to\mu')g(\mu') \qquad (4)$$

where λ is the extinction coefficient (least eigenvalue of eq(1)) and g(μ) the asymptotic angular intensity distribution. The K-integrals obey diffusion-type equations, which are however exact since they are derived from the radiative transfer equation

### 2.2 Isotropic scattering

For isotropic scattering, f(μ' → μ) ≡ ½ and

$$g(\mu)=\frac{\varpi}{2(1-\lambda\mu)} \qquad (5)$$

The eigenvalue (extinction coefficient) λ is given by the normalisation

$$1=\frac{\varpi}{2\lambda}\log\left(\frac{1+\lambda}{1-\lambda}\right) \qquad (6)$$

and the corresponding weight functions [4]

$$g_1(\mu)=\frac{\varpi}{1-\lambda^2\mu^2} \qquad (7a)$$

$$g_2(\mu)=\frac{\varpi\mu}{1-\lambda^2\mu^2} \qquad (7b)$$

The corresponding formulae for anisotropic scattering are discussed in Appendix A.

**2.3 Fresnel boundary equations**

For a half-space with dielectric boundary at x = 0, refractive index n>1 (x>0), the intensity in the forward hemisphere I(0, μ) (μ>0) comprises the reflected part of the intensity in the backward hemisphere I(0,− μ) plus the contribution of the surface source $S_0(\mu_0)$ [4]

$$I(0,\mu) = R(\mu,\mu_0)I(0,-\mu) + n^2[1 - R(\mu,\mu_0)]S_0(\mu_0) \qquad (\mu \geq 0) \qquad (8)$$

where R(μ, $\mu_0$) is the Fresnel reflectance for unpolarised light [2]

$$R(\mu) = \frac{1}{2}\left[\left(\frac{\mu - n\mu_0}{\mu + n\mu_0}\right)^2 + \left(\frac{\mu_0 - n\mu}{\mu_0 + n\mu}\right)^2\right] \qquad \mu \geq \mu_c \qquad (9)$$

where $\mu_0$ is the cosine of the angle of incidence, μ the cosine of the angle of refraction and $\mu_c$ the cosine of the critical angle for internal reflection: $\mu_c^2 = 1 - 1/n^2$. The factor $n^2$ in eqn (8) arises from refraction at the boundary, where light filling a hemisphere on the lower index side is refracted into a cone of semi-vertex angle $\cos^{-1}\mu_c$ on the higher index side and vice-versa. For a dielectric layer of finite width d (slab), the intensity incident at the second boundary I(d, μ) is partially reflected, thus

$$I(d, -\mu) = R(\mu)I(d, \mu) \qquad (10)$$

At an interface between two dielectric media with a boundary at x = 0, the forward intensity I(x, μ') for μ' >0, x ≥ 0 comprises the fraction of the backward intensity I(x, -μ') reflected at the boundary plus the forward intensity transmitted at the boundary from the adjoining medium viz.

$$I(0, \mu') = R(\mu, \mu')I(0, -\mu') + n^2[1 - R(\mu, \mu')]I(0, \mu) \qquad (11)$$

where n>1 is the ratio of the refractive indices and μ, μ' are related by

$$1 - \mu^2 = n^2(1 - \mu'^2) \qquad (11')$$

from Snell's law. Similarly, the backward intensity I(x, -μ) at the boundary in the adjoining medium for μ>0, x ≤ 0. comprises the fraction of the forward intensity I(x, μ) reflected at the boundary plus the backward intensity transmitted by the boundary viz.

$$I(0, -\mu) = R(\mu, \mu')I(0, \mu) + [1 - R(\mu, \mu')]I(0, -\mu')/n^2 \qquad (12)$$

**2.4 Boundary transients**

In planar geometry with rotational symmetry, the asymptotic radiant intensity I(x, μ) in the depth of a turbid medium far from the boundary may be written

$$I(x, \mu) = A\exp(-\lambda x)g(\mu) \qquad (13)$$

where the x-axis is normal to the plane of symmetry, λ is the extinction coefficient and μ the cosine of the angle between I(x, μ) and the symmetry axis; g(μ) is the asymptotic angular intensity distribution (Sec 2.1 above). The intensity near a boundary is obtained by adding a boundary transient ψ(x, μ), which decays exponentially away from the boundary. Thus for a half-space with boundary at x = 0 [4]

$$I(0, \mu) = Ag(\mu) + B\psi(0, \mu) \qquad (14)$$

Similarly, the intensities at the boundaries x = ± a of a finite layer (slab) of width d = 2a may be expressed as the sums of the asymptotic terms and the boundary transients $\psi_1(-a, \mu)$, $\psi_2(a, \mu)$

$$I(-a, \mu) = A_1 \exp(\lambda a) g(\mu) + A_2 \exp(-\lambda a) g(-\mu) + B_1 \psi_1(-a, \mu) \qquad x = -a \qquad (15a)$$

$$I(a, \mu) = A_1 \exp(-\lambda a) g(\mu) + A_2 \exp(\lambda a) g(-\mu) + B_2 \psi_2(a, \mu) \qquad x = a \qquad (15b)$$

Before the calculation can proceed it is necessary to select suitable approximations to the unknown boundary transients $\psi(x, \mu)$, $\psi_1(-a, \mu)$, $\psi_2(a, \mu)$ [3, 4]. A simple choice is to set $\psi(\mu) = 1$ for initial evaluation of the coefficients B, $B_1$, $B_2$; a more informed choice for the boundary transient (for isotropic scattering) follows from [9]

$$\psi(x, \mu) = \int_0^1 dw A(w) \varphi(w, \mu) e^{-x/w} \qquad (16)$$

where

$$\varphi(w, \mu) = \frac{\varpi\, w}{2(w - \mu)} + [1 - \frac{\varpi\, w}{2} \ln\left(\frac{1+w}{1-w}\right)] \delta(w - \mu) \qquad (17)$$

and A(w) is undetermined. Setting A(w) = $A_1$ (constant) we find for x = 0 and $\mu \rightarrow -\mu$,

$$\psi(0, -\mu) \approx A_1 \int_0^1 dw \varphi(w, -\mu) \qquad (18)$$

whence

$$\psi(0, -\mu) \approx A_1 \frac{\varpi}{2} (1 - \mu \ln(1 + 1/\mu)) \qquad (19)$$

while for a slab of finite width d = 2a the transient will include an exponential term ~exp(-2a/μ). This form of boundary transient can produce remarkably accurate values for the boundary fluxes φ and **J** and half-space/slab albedo (diffuse reflectance) for isotropic scattering, particularly in the limits (ϖ, n) ⇒ 1 (see below). Further increases in accuracy can be achieved by series expansion of A(w) in eqn (16) above and consequent improvement in ψ(0, μ), requiring higher moments of the transfer equation in addition to the K-integrals [10]

Alternatively, we may employ a simple one-parameter linear or quadratic form for the unknown boundary transient viz.

$$\psi(0, \mu) = 1 + \beta \mu^k \qquad k = 1, 2 \qquad (19')$$

requiring an extra moment of the RTE at a boundary to evaluate β . Where accurate slab albedo and transmittance data are available e.g. [11], the value of β in (19') can be determined and the matching boundary intensity I(0, μ, β ) compared with that following from eqn (19).

### 2.5 Integrated boundary equations

Multiplying the Fresnel boundary eqn (8) through by $\mu\, g_i(\mu)\, d\mu$ and integrating over the interval $\mu\in[0, 1]$, we obtain the integrated boundary equations (i =1, 2) for the half-space

$$A_1\int_0^1 d\mu\mu g_i(\mu)[g(\mu)-R(\mu)g(-\mu)]+B_1\int_0^1 d\mu\mu g_i(\mu)[\psi(0,\mu)-R(\mu)\psi(0,-\mu)]$$
$$=n^2\int_0^1 d\mu\mu g_i(\mu)[1-R(\mu,\mu_0)]S(\mu,\mu_0) \qquad (20)$$

where $\psi(0, \mu)$ can be replaced by $\psi(0, -\mu)$ in the integrand using the result [4]

$$\int_0^1 d\mu\mu g_i(\mu)\psi(0,\mu)=\int_0^1 d\mu\mu g_i(-\mu)\psi(0,-\mu) \qquad (21)$$

which follows from the vanishing of the K-integral of the boundary transient

$$\int_{-1}^1 d\mu\mu g_i(\mu)\psi(x,\mu)=0 \qquad (22)$$

Setting i =1 and i = 2 in turn in eqn (20) yields two simultaneous equations for $A_1$ and $B_1$ which can be written in matrix form as

$$\begin{vmatrix} I_1 & \psi_1 \\ I_2 & \psi_2 \end{vmatrix} \begin{vmatrix} A_1 \\ B_1 \end{vmatrix} = n^2 \begin{vmatrix} S_1 \\ S_2 \end{vmatrix} \qquad (23)$$

and solved for $A_1$ and $B_1$ by matrix inversion, where $I_i$ ,$\Psi_i$ and $S_i$ are the K-integrals defined in (20).

The outward intensity $I(0, -\mu)$ at the boundary follows from eqn (14) and the emergent vector flux

$$J(0)=-\frac{1}{n^2}\int_0^1 d\mu\,\mu\, I(0,-\mu)(1-R(\mu)) \qquad (24)$$

enabling the diffuse reflectance (albedo) to be calculated.

Similarly, two sets of integrated boundary equations for the slab follow from eqns (15), yielding four simultaneous equations for $A_1$, $A_2$, $B_1$, $B_2$. These can be written in matrix form as

$$\mathbf{MA=S} \qquad \begin{vmatrix} m_{11} & m_{12} & m_{13} & m_{14} \\ m_{21} & m_{22} & m_{23} & m_{24} \\ m_{12} & m_{11} & m_{14} & m_{13} \\ m_{22} & m_{21} & m_{24} & m_{23} \end{vmatrix} \begin{vmatrix} A_1 \\ A_2 \\ B_1 \\ B_2 \end{vmatrix} = n^2 \begin{vmatrix} S_1 \\ S_2 \\ Q_1 \\ Q_2 \end{vmatrix} \qquad (25)$$

**M** is a 4x4 matrix with pairs of elements in the first two rows reversed in rows 3 and 4, $S_1$, $S_2$ are the source integrals for the first surface, $Q_1$, $Q_2$ those for the second surface ($Q_1 = Q_2 \equiv 0$ if no light is externally incident on the second surface). The coefficients $A_1$, $A_2$, $B_1$, $B_2$ are found by matrix inversion viz. $\mathbf{A = M^{-1}S}$, enabling the boundary intensities $I(a,-\mu)$ and $I(a, \mu)$ to be determined from eqns (15a, b), from which the emergent vector fluxes are calculated as

$$J(-a)=-\frac{1}{n^2}\int_0^1 d\mu\mu I(-a,-\mu)(1-R(\mu)) \qquad (26a)$$

$$J(a)=\frac{1}{n^2}\int_0^1 d\mu\mu I(a,\mu)(1-R(\mu)) \quad (26b)$$

The diffuse reflectance R* and transmittance T* follow on dividing J(–a), J(a) by the incident flux

$$F_0=\int_0^1 d\mu_0\mu_0 S_0(\mu_0) \quad (27)$$

Thus the slab albedo

$$A^*=R^*+R_{ext}$$

where

$$R_{ext}=\int_0^1 d\mu_0\mu_0 R(\mu_0)S(\mu_0) \quad (28)$$

is the external reflectance of the dielectric surface averaged over the surface source function $S_0(\mu_0)$. The limiting cases of perfect and zero scattering ($\varpi$ =1, 0) are discussed in Appendix III.

The integrated boundary equations for the interface between two adjoining scattering media with differing refractive index, obtained from the K-integrals of eqns (11) and (12), likewise yield a 4x4 matrix for the coefficients, while those for a double layer, comprising an interface and two exterior boundaries, yield an 8x8 matrix. The addition of further layers can be accommodated by further expanding the matrix **M** and solving by matrix inversion.

## 3. Results

Examples of K-integral calculations for isotropic scattering in slab and half-space geometries with diffuse and planar illumination are compared with radiative transfer in the following Tables and Figures. Results for slab albedo A* (diffuse reflectance) for n = 1, d = 1, 5, 10 and $\varpi$ = 0.7, 0.9, 1 are presented in Table Ia for planar illumination at non-normal incidence ($\mu^*$ = 0.9) and for isotropic illumination respectively, the accuracy of the K-integral results increasing with increasing slab thickness d and single scattering albedo c. Thus 5-figure accuracy is attained for $\varpi$ = 1, d = 10, while for $\varpi$ = 0.7, d = 1 the error is 0.7% for planar and 3% for isotropic illumination. Results for a half-space (n = 1) and $\varpi$ = 0.7, 0.9, 0.99 and 0.999 are compared with radiative transfer data in Table Ib, showing similar accuracy [12, 13]. K-integral results for albedo A* and diffuse transmittance T* for isotropic illumination of slabs of thickness d = 1, 2, 5, 10 for a wide range of single scattering albedoes ($\varpi \in |0.2, 1|$) are compared with radiative transfer data in Table II for n = 1, the accuracy again increasing with $\varpi$ and d, ranging from 5-figure accuracy for $\varpi$ = 1, d = 10 to 1.5% error in A* for $\varpi$ = 0.2, d = 2 [11-14]. K-integral results for the diffuse reflectance (albedo) A* and diffuse transmittance T* of slabs of thickness d = 10, 1 and refractive index n = 1.5 are compared with radiative transfer data in Tables IIIa, b for isotropic surface illumination. Again it can be seen that K-integral calculations of slab albedo are remarkably accurate, agreeing with radiative transfer within 2 units in the 5th decimal place for c = 0.3, 0.7 and 1 for d = 10, though less accurate for d = 1 [11]. K-integral results for albedo A* and diffuse transmittance T* of double layers, comprising two adjacent slabs of differing refractive index, requiring an 8x8 matrix solution (Sec 2.5), are compared with RT data in Tables IVa, b, c. Best results yield errors of only 1-2 units in the 5th decimal place, as for the single slab [11]. Table V shows the dependence of albedo A* on incidence angle for plane-wave illumination of a half-space, comparing the results of diffusion and K-integral calculations with RT data [15], the percentage error in the K-integral results being ~100 times smaller than the error in the diffusion values. K-integral calculations of albedo A* and transmittance T* for obliquely incident planar illumination of refracting slabs are compared with radiative transfer data in Table VI, again showing excellent agreement for d = 10, but are less accurate for d = 1, 0.1 [16]. The supremacy of the K-integral approach over diffusion theory is illustrated in Table VII, where detailed comparison of diffusion theory with the asymptotic (Appendix B) and K-integral results for slab albedo A* and diffuse transmission T* for isotropic illiumination is presented [11]. It can be seen that the diffusion error is substantial for $\varpi < 0.99$, exceeding 10% for $\varpi<0.9$, though achieving near 4-figure accuracy for $\varpi = 1$ (perfect scattering), while the error in the asymptotic solution is ~2% for $\varpi = 0.7$, decreasing to 0.4% when $\varpi = 1$. The error in the K-integral result is orders of magnitude smaller viz. $\delta = 3\times10^{-4}$ for $\varpi = 1$, $1\times10^{-4}$ for $\varpi = 0.7$, $2\times10^{-4}$ for $\varpi = 0.3$. The effect of the choice of boundary transient on the K-integral values of A* and T* is shown in Table VIII. The first three columns give the asymptotic, constant term and log function (eqn (19)) results for n = 1.5 and slab thicknesses L = 10, 1 compared with the RT values (last column), the accuracy generally improving with improved choice of transient term. The subsequent columns show the exact fit enabled by the introduction of an adjustable parameter in a simple linear $(1-\beta\mu)$ or quadratic transient $(1-\alpha\mu^2)$.

The parametric dependence of the K-integral solutions are shown in the following Figures: Fig. 1 shows the dependence of the diffuse reflectance (albedo A*) on particle scattering albedo for isotropic surface illumination of a half-space (n = 1) with isotropic scattering. The K-integral results (continuous curve) show excellent agreement with the radiative transfer (RT) data points (diamonds) [17] for the full range of particle scattering albedo$\varpi \in [0, 1]$ . Fig. 2 shows the excellent agreement of K-integral calculations of the albedo A* with RT data for planar illumination of a half-space for two values of refractive index (n = 1.4, 1.6) [18]. Fig 3 shows the excellent agreement of K-integral calculations of the albedo A* with RT data for planar illumination of a slab (d=5, n=1.4) for the full range of particle scattering albedoes $\varpi \in |0, 1|$. Diffusion theory rapidly diverges from the accurate results as scattering albedo diminishes, while the modified 2-flux theory of [18] shows much better agreement. Fig 4 shows excellent agreement of the K-integral results (continuous curve) with radiative transfer data (circles) for the full range of scattering albedoes $\varpi \in |0, 1|$ for the diffuse reflectance of a slab under isotropic illumination [11]; it also shows the rapid divergence between radiative transfer and diffusion theory for particle scattering albedo $\varpi < 0.9$. The dependence of the diffuse reflectance (albedo A*) on scattering albedo $\varpi$ for planar illumination of a half-space is compared for δ-isotropic and Henyey-Greenstein scattering in Fig 5, again showing excellent agreement with RT data [18]. The dependence of the diffuse reflectance of a half-space on refractive index for a range of particle scattering albedoes is presented in Fig 6, showing excellent agreement with the RT data [15]; diffusion theory (dashed curves) is clearly less accurate than the K-integral results, particularly as $n \Rightarrow 1$. The dependence on incident cosine $\mu^*$ of the diffuse reflectance and transmittance for planar illumination of a dielectric slab backed by an index matched non-scattering half-space are shown in Fig 7, comparing K-integral calculations with RT data for slabs of width L = 10, 1, 0.1 [16]. The K-integral values show increasing divergence from the RT data for thinner layers as $\mu^* \Rightarrow 1$ (normal incidence). The dependence of diffuse reflectance (albedo A*) on incident angle for plane-wave illumination of a half-space is shown in Fig. 8, comparing the K-integral results with diffusion theory and radiative transfer data for $\varpi = 0.99$ [15]. Figs 9-11 show the angular intensity profiles emerging at the entry and exit surfaces of a finite scattering layer, comparing K-integral results with radiative transfer data [12, 13]. Fig. 9 shows the intensity profile exiting the far side of a scattering layer of width d = 5 for isotropic incident light, Fig. 10 shows the intensity profile of the backscattered light at the incident surface, Fig. 11 shows the backscattered profile for oblique plane-wave illumination of a d = 5 scattering layer. Here the agreement is less good for angles $\theta > 70$ deg but still within ~2% at $\theta = 87$ deg ($\mu = 0.05$). The angular emission intensities exiting the front and back surfaces of a dielectric slab (n= 1.4) are shown in Fig 12, the emitted intensities diminishing with increasing exit angle θ as the internal incidence approaches the citical angle for totsal internal reflection. Fig. 13 shows the dependence of slab albedo A*, diffuse transmittance T* and diffuse flux vector J- at the illuminated surface on refractive index n. K-integral calculations of the difference in scalar flux on either side of an interface (refractive index discontinuity) are compared with diffusion and asymptotic theory in Fig. 14, the asymptotic results having a considerably smaller error [4]

## 4. Summary and conclusions

The examples presented here show how accurate solutions for radiative transfer in scattering media with Fresnel reflection at the boundaries can be obtained by the method of K-integrals, originally developed for solving boundary problems in neutron transport theory [7, 8]. Agreement to better than 0.0025% with radiative transfer solutions is achieved for the diffuse reflectance (albedo) and diffuse transmittance of slabs with dielectric boundaries (Table III). For slabs with free boundaries (n=1), five-figure accuracy is obtained in some cases (Tables I, II). Accurate results for diffuse isotropic and plane wave illumination (normal and oblique incidence) of dielectric layers are achieved for particle scattering albedoes as low as $\varpi = 0.2$ and slab thickness d = 1 mfp, a far greater range than is available with diffusion solutions (Tables III, VI, VII). The angular emission intensities also show excellent agreement with exact radiative transfer, even though only their weighted means are used in the K-integrals (Figs 9, 10, 11). The key to this success lies in the fact that the K-integrals are independent of the boundary conditions, though they are solutions of the radiative transfer equation [4, 7, 8]. By requiring an analytic solution, comprising the asymptotic intensity valid in the depth of the turbid medium plus a boundary transient, to satisfy the K-integrals, we obtain integrated boundary equations providing highly accurate solutions to the problems considered here. The use of integrated boundary equations greatly simplifies the calculations, requiring just two simultaneous equations for evaluating the relative intensities of the transient and asymptotic terms for a half-space (four for a slab or an interface), rather than the large number of terms needed for similar accuracy with PN expansions, the coefficients being averaged over the angular intensity distribution. The results can be highly accurate when compared with radiative transfer theory [4, 11]. The principal advantage of the K-integral method lies in its straightforward application to some challenging problems in radiative transfer not readily tackled by diffusion theory, without the need to solve the radiative transfer equation (RTE) directly

**Acknowledgment**

I wish to express my heartfelt gratitude to Mike Williams for introducing me to the K-integral method of solving transport problems involving Fresnel reflection at the boundaries, for his unfailing readiness to respond to my endless queries and for supplying numerous reprints of relevant publications from his extensive archive of transport literature.

**Appendix A**

**Anisotropic scattering**

For forward-biased anisotropic scattering, the transport approximation [19], comprising a delta-function added to the isotropic scattering term

$$f(\mu,\mu') = 2\beta\delta\,|\,\mu-\mu'| + 1-\beta \qquad 0<\beta<1 \qquad \text{(A1)}$$

where β is the forward scattered fraction, provides a simple approach to anisotropic scattering via a rescaling of the parameters, namely optical depth τ and scattering albedo ϖ

$$\tau' = \tau(1-\beta\varpi) \qquad \text{(A2)}$$

$$\varpi' = \varpi\frac{1-\beta}{1-\beta\varpi} \qquad \text{(A3)}$$

Calculations are made with the rescaled parameters assuming isotropic scattering and can yield good results for comparison with realistic forward-biased scattering functions. To proceed further e.g. with the δ-P1 approximation, which includes a linear term [20]

$$f(\mu,\mu') = 2\beta\delta\,|\,\mu-\mu'| + (1-\beta)(1+3b\mu\mu') \qquad 0<\beta<1 \qquad \text{(A4)}$$

the transfer equation can be approximated by a rescaled diffusion equation [21], or for accurate results with more realistic scattering functions, such as the Henyey-Greenstein phase function [22], the asymptotic intensity distribution g(μ) may be calculated from eq (4) for a specific phase function f(μ, μ') [23] and inserted in the K-integral eqns for solution. Thus expanding g(μ) as a series of Legendre polynomials $P_n(\mu)$ [24]

$$g(\mu) = \sum_{n=0}^{\infty}(2n+1)b_n P_n(\mu) \qquad \text{(A5)}$$

inserting in eq (4) and equating terms we obtain the recursion relations for the coefficients $b_n$

$$b_0 - \gamma b_1 = \varpi b_0 g_0 \qquad \text{(A6)}$$

$$\lambda(n+1)b_{n+1} - (2n+1)(1-\varpi g_n)b_n + \lambda n b_{n-1} = 0 \qquad \text{(A7)}$$

where the $g_n$ are the coefficients of the Legendre polyniomial series for the phase function

$$p(\mu,\mu') = \sum_m (2m+1) g_m P_m(\mu) P_m(\mu') \qquad \text{(A8)}$$

and λ is the extinction coefficient (least eigenvalue of the characteristic equation (eqn (4))

λ can be calculated to any desired degree of accuracy from the recursion relation [25]

$$\Delta_n = \frac{(n\lambda)^2}{(2n+1)(1-\varpi g_n) - \Delta_{n+1}} \qquad \text{(A9)}$$

with $$1-\varpi = \Delta_1 \qquad \text{(A10)}$$

Alternatively, we may employ the analysis of Haltrin [25], who derived a phase function

$$f(\mu, \mu') = 2\beta\delta\,|\,\mu - \mu'| + (1-\beta)h(\mu\mu') \qquad 0 < \beta < 1$$

$$h(\mu, \mu') = \sum_m P_m(\mu)P_m(\mu') \tag{A11}$$

assuming a Henyey-Greenstein asymptotic radiance distribution

$$g(\mu) = HG(\eta, \mu) = \frac{1-\eta^2}{(1-2\eta\mu+\eta^2)^{3/2}} \tag{A12}$$

valid in the depth of the scattering medium. This has the merit of retaining simple analytic formulae for the weight-functions $g_i(\mu)$. The mean cosine $\eta$ of the asymptotic radiance is determined by the single scattering albedo $\varpi$ and the forward scattered fraction $\beta$; when $\beta = 0$

$$\eta^2 = \frac{1-\varpi}{1+\varpi} \tag{A13}$$

The eigenvalue then follows from the Gershun relation [26]

$$\lambda = \frac{1-\varpi}{\eta} = \sqrt{1-\varpi^2} \tag{A14}$$

The corresponding weight-functions $g_i(\mu)$ are (from eqns (3a, b) and eqn (31))

$$g_1(\mu) = \frac{1}{2}[HG(\eta, \mu) + HG(\eta, -\mu] \tag{A15}$$

$$g_2(\mu) = \frac{1}{2\lambda}[HG(\eta, \mu) - HG(\eta, -\mu)] \tag{A16}$$

These can then be applied in the analysis for comparison with e.g. the transport approximation or the δ-Eddington approximation for anisotropic scattering

**Appendix B**

**Asymptotic theory**

The asymptotic approach employs the partial currents $J_+$, $J_-$ obtained by integrating the vector component of the asymptotic intensity $I(\mu)$ over the forward and backward hemispheres [27]. In contrast to the δ-P1 diffusion analysis, which assumes a simple cosine intensity variation [21], the accurate asymptotic intensity distribution corresponding to the scattering phase function is used here, while the boundary transient is omitted. The asymptotic calculations yield more accurate results than diffusion theory, they halve the number of boundary equations (compared with the K-integrals), thereby reducing the numerical effort in multilayer calculations, and are applicable over a broad range of scattering albedoes $\varpi \in [0, 1]$

The partial currents $J_+$, $J_-$ are defined as follows

$$J_+ = ½ \int_0^1 I(\mu)\mu d\mu \qquad \text{(B1)}$$

$$J_- = ½ \int_0^1 I(-\mu)\mu d\mu \qquad \text{(B2)}$$

At a dielectric boundary

$$I(\mu) = R(\mu) I(-\mu) \qquad (\mu > 0) \qquad \text{(B3)}$$

where R(μ) is the Fresnel reflectance function (eqn()). Multiplying by μ dμ and integrating yields

$$J_+ = ½ \int_0^1 I(\mu)\mu d\mu = ½ \int_0^1 R(\mu) I(-\mu)\mu d\mu \qquad \text{(B4)}$$

Thus the asymptotic IBE for a half-space (cf eqn (20)) is

$$I_0 \int_0^1 [g(\mu) - R(\mu) g(-\mu)]\mu\, d\mu = n^2 \int_0^1 [1 - R(\mu)] S(\mu)\mu d\mu \qquad \text{(B5)}$$

where $I_0$ g(μ) is the asymptotic intensity distribution and S(μ) the surface source

For isotropic scattering

$$g(\mu) = \frac{\varpi}{2(1-\lambda\mu)} \qquad \text{(B6)}$$

and the IBE for the half-space becomes

$$A_0 \int_0^1 [\frac{1}{1-\lambda\mu} - \frac{R(\mu)}{1+\lambda\mu}]\mu\, d\mu = n^2 \int_0^1 [1 - R(\mu)] S(\mu)\mu\, d\mu \qquad \text{(B7)}$$

enabling the intensity $A_0$ to be evaluated directly in terms of the source strength. The first term is analytic, the terms with R(μ) are evaluated by numerical integration, and can be tabulated for isotropic scattering and an analytic source function e.g. a Lambertian (cosine) surface source.

For a single layer (slab) of width d = 2a, the asymptotic intensities at the boundaries are

$$I(a, \mu) = A_1 \exp(-\lambda a)\, g(\mu) + A_2 \exp(\lambda a)\, g(-\mu) \qquad x = a \qquad \text{(B8a)}$$

$$I(-a, \mu) = A_1 \exp(\lambda a)\, g(\mu) + A_2 \exp(-\lambda a)\, g(-\mu) \qquad x = -a \qquad \text{(B8b)}$$

And the IBEs for a surface source S(μ) at x = -a and isotropic scattering are

$$A_1 \exp(-\lambda a) \int_0^1 [\frac{1}{1+\lambda\mu} - \frac{R(\mu)}{1-\lambda\mu}]\mu\, d\mu + A_2 \exp(\lambda a) \int_0^1 [\frac{1}{1-\lambda\mu} - \frac{R(\mu)}{1+\lambda\mu}]\mu\, d\mu = 0 \qquad (x = a)$$

$$A_1 \exp(\lambda a) \int_0^1 [\frac{1}{1-\lambda\mu} - \frac{R(\mu)}{1+\lambda\mu}]\mu\, d\mu + A_2 \exp(-\lambda a) \int_0^1 [\frac{1}{1+\lambda\mu} - \frac{R(\mu)}{1-\lambda\mu}]\mu\, d\mu = 0$$

$$= n^2 \int_0^1 [1 - R(\mu)] S(\mu)\mu\, d\mu \qquad (x = -a) \qquad \text{(B9a, b)}$$

enabling the coefficients $A_1$ and $A_2$ to be evaluated. Results for the asymptotic calculation of albedo and transmittance of a slab are compared with the diffusion,K-integral and RT values in Table VII. It can be seen that the asymptotic values are much closer to the K-integral and RT data than the diffusion values.

**Appendix C**

**Scattering limits**

**1. Perfect scattering**

In the limit of perfect scattering ($\varpi = 1$), the extinction coefficient vanishes ($\lambda = 0$) and the asymptotic intensity I(μ) is isotropic: g(μ) = 1 for μ ∈ |−1, 1|. Hence $g_1(\mu) = 1$, $g_2(\mu) = \mu$ and the integrated boundary equation (IBE) for the half-space (eqn (20)) reduces to

$$A_1\int_0^1 d\mu\mu^i[1-R(\mu)] + B_1\int_0^1 d\mu\mu^i[\psi(0,\mu)-R(\mu)\psi(0,-\mu)]$$

$$= n^2\int_0^1 d\mu\mu^i[1-R(\mu,\mu_0)]S(\mu,\mu_0) \qquad i=1,2 \qquad \text{(C1)}$$

with similar IBEs for a slab (from eqns (15a, b)). Eqn (29) contains analytic integrals of the form

$$J_k = \int_0^1 [1-R(\mu)]\mu^k d\mu \qquad \text{(C2)}$$

as discussed by Aronson [ref]; they can also be evaluated numerically. The source integral (rhs of eqn (29)) takes the same form for isotropic illumination viz. $S(\mu, \mu_0) = S_0$; only the K-integral of the product $R(\mu)\psi(0, -\mu)$ may be non-analytic and require numerical evaluation.

**2. Zero scattering**

In the zero scattering limit $\varpi = 0$, $\lambda = 1$ and the K-integrals diverge; light is reflected and refracted at the boundaries and absorbed in the medium, the albedo and transmittance being determined by Fresnel reflectance and optical depth. For a thick slab or half-space, the albedo is given by the surface reflectance $R_{ext}$ averaged over the angular intensity distribution of the source (eqn (28)). For a finite slab, the contribution of internal absorption and multiple reflection at the boundaries is taken into account. This can be done without approximation, the calculations being exact, even for thin slabs.

For a surface source $S(\mu_0)$, the slab albedo A* and transmittance T* of a slab of width 2a are

$$A^* = R_{ext} + \int_0^1 d\mu_0\mu_0[1-R(\mu_0)]S(\mu_0)R(\mu)\frac{[1-R(\mu)]e^{-4a/\mu}}{1-R^2(\mu)e^{-4a/\mu}} \qquad \text{(C3)}$$

$$T^* = \int_0^1 d\mu_0\mu_0[1-R(\mu_0)]S(\mu_0)\frac{[1-R(\mu)]e^{-2a/\mu}}{1-R^2(\mu)e^{-4a/\mu}} \qquad \text{(C4)}$$

in the zero scattering limit.

| | n = 1 | | | |
|---|---|---|---|---|
| | planar | δ\|0.9–μ\| | isotropic | |
| | K-integral | Siewert | K-integral | Siewert |
| | ϖ = 1 | | ϖ = 1 | |
| d | A* | A* | A* | A* |
| 1 | 0.34854 | 0.365087 | 0.449103 | 0.446594 |
| 5 | 0.754448 | 0.754496 | 0.792325 | 0.792343 |
| 10 | 0.861949 | 0.861963 | 0.883249 | 0.883255 |
| | ϖ = 0.9 | | ϖ = 0.9 | |
| 1 | 0.2762 | 0.286298 | 0.357297 | 0.352712 |
| 5 | 0.428115 | 0.428393 | 0.476225 | 0.476338 |
| 10 | 0.430255 | 0.43053 | 0.477901 | 0.478016 |
| | ϖ = 0.7 | | ϖ = 0.7 | |
| 1 | 0.179285 | 0.178039 | 0.22922 | 0.22207 |
| 5 | 0.219589 | 0.219464 | 0.25638 | 0.256519 |
| 10 | 0.219642 | 0.219519 | 0.25642 | 0.256557 |

Table Ia

| | half-space | | half-space | |
|---|---|---|---|---|
| | δ\|0.9–μ\| | | isotropic | |
| ϖ | K-integral | Siewert | K-integral | Siewert |
| | A* | A* | A* | A* |
| 0.7 | 0.219642 | 0.219519 | 0.25642 | 0.256557 |
| 0.9 | 0.430266 | 0.430541 | 0.47791 | 0.478025 |
| 0.99 | 0.764234 | 0.764306 | 0.79455 | 0.794564 |
| 0.999 | 0.917707 | 0.91773 | 0.929709 | 0.929713 |

Table Ib

Diffuse reflectance (albedo A*) of plane layers (Table Ia) and half-spaces (Table Ib) for oblique planar and isotropic illumination (n=1). Garcia&Siewert [12, 13]

| n = 1 | d = 1 | Wu, Liou | | Wu, Liou |
|---|---|---|---|---|
| ϖ | isotropic | Siewert | isotropic | Siewert |
| | K-integral | Degheidy | K-integral | Degheidy |
| | A* | A* | T* | T* |
| 0.2 | 0.04682 | 0.04393 | 0.0046 | 0.24627 |
| 0.3 | 0.0752 | 0.0701 | 0.0476 | 0.26269 |
| 0.4 | 0.1067 | 0.0999 | 0.1245 | 0.28254 |
| 0.5 | 0.1419 | 0.1342 | 0.2033 | 0.30663 |
| 0.6 | 0.18207 | 0.1743 | 0.2732 | 0.33524 |
| 0.7 | 0.22922 | 0.22207 | 0.3366 | 0.371195 |
| 0.8 | 0.28623 | 0.28015 | 0.399 | 0.41624 |
| 0.9 | 0.357297 | 0.352712 | 0.4675 | 0.474746 |
| 0.995 | 0.44387 | 0.44124 | 0.5462 | 0.54884 |
| 1 | 0.449092 | 0.446594 | 0.5509 | 0.553406 |

| | d = 2 | | | |
|---|---|---|---|---|
| ϖ | isotropic | | isotropic | |
| | K-integral | Degheidy | K-integral | Degheidy |
| | A* | A* | T* | T* |
| 0.2 | 0.0468 | 0.0461 | 0.0017 | 0.0728 |
| 0.3 | 0.0753 | 0.0741 | 0.0176 | 0.0814 |
| 0.4 | 0.1078 | 0.1066 | 0.0465 | 0.0925 |
| 0.5 | 0.1461 | 0.1451 | 0.078 | 0.1071 |
| 0.6 | 0.1927 | 0.1919 | 0.1101 | 0.1269 |
| 0.7 | 0.2513 | 0.2506 | 0.1462 | 0.1551 |
| 0.8 | 0.3285 | 0.328 | 0.1932 | 0.1973 |
| 0.9 | 0.4376 | 0.4376 | 0.2642 | 0.2656 |
| 1 | 0.6102 | | 0.3898 | |

| | d = 5 | | | |
|---|---|---|---|---|
| ϖ | isotropic | Siewert | isotropic | Siewert |
| | K-integral | Degheidy | K-integral | Degheidy |
| | A* | A* | T* | T* |
| 0.2 | 0.0468 | 0.0463 | 0.000084 | 0.0024 |
| 0.3 | 0.0753 | 0.0745 | 0.00088 | 0.003 |
| 0.4 | 0.108 | 0.1073 | 0.0024 | 0.0039 |
| 0.5 | 0.14687 | 0.1465 | 0.0044 | 0.0053 |
| 0.6 | 0.1947 | 0.1947 | 0.0072 | 0.0077 |
| 0.7 | 0.25638 | 0.25652 | 0.012154 | 0.012389 |
| 0.8 | 0.3415 | 0.3417 | 0.0228 | 0.0229 |
| 0.9 | 0.47623 | 0.47634 | 0.053405 | 0.053421 |
| 1 | 0.792353 | 0.792343 | 0.207645 | 0.207657 |

| | d = 10 | | | |
|---|---|---|---|---|
| ϖ | isotropic | Wu, Liou | isotropic | Wu, Liou |
| | K-integral | Siewert | K-integral | Siewert |
| | A* | A* | T* | T* |
| 0.2 | 0.0468 | 0.04626 | 5.7E-007 | 0.00001 |
| 0.3 | 0.0753 | | 0.000006 | |
| 0.4 | 0.108 | | 0.000017 | |
| 0.5 | 0.1469 | | 0.000037 | |
| 0.6 | 0.1947 | | 0.000077 | |
| 0.7 | 0.2564 | 0.256557 | 0.000193 | 0.000194 |
| 0.8 | 0.3417 | 0.34187 | 0.00065 | 0.00065 |
| 0.9 | 0.4779 | 0.478016 | 0.0039 | 0.003856 |
| 0.95 | 0.5964 | | 0.0141 | |
| 0.995 | 0.8284 | 0.82841 | 0.08667 | 0.08667 |
| 1 | 0.883249 | 0.883255 | 0.116751 | 0.116745 |

Table II

Albedo A* and diffuse transmittance T* of plane layers (n=1) for isotropic surface illumination (Degheidy, Garcia&Siewert, Wu&Liou [11-14])

n = 1.5 d = 10
isotropic incident light, isotropic scattering

| ϖ | A* K-integral | A* Wu, Liou | T* K-integral | T* Wu, Liou |
|---|---|---|---|---|
| 0.1 | 0.09226 | | 3E-010 | |
| 0.2 | 0.10428 | | 1.00E-06 | |
| 0.3 | 0.11247 | 0.11245 | 9.47E-06 | 0.00002 |
| 0.4 | 0.12271 | | 0.000024 | |
| 0.5 | 0.13603 | | 0.000044 | |
| 0.6 | 0.15413 | | 0.000081 | |
| 0.7 | 0.18036 | 0.18034 | 0.00018 | 0.00018 |
| 0.8 | 0.22268 | | 0.00057 | |
| 0.9 | 0.30803 | | 0.003545 | |
| 0.95 | 0.40493 | | 0.01479 | |
| 0.975 | 0.50514 | | 0.04107 | |
| 0.99 | 0.62461 | | 0.09487 | |
| 0.995 | 0.69244 | | 0.13507 | |
| 0.999 | 0.76994 | | 0.18766 | |
| 1 | 0.79447 | 0.79445 | 0.20553 | 0.20555 |

Table III a

n = 1.5 d = 1
isotropic incident light, isotropic scattering

| ϖ | A* K-integral | A* Wu, Liou | T* K-integral | T* Wu, Liou |
|---|---|---|---|---|
| 0.3 | 0.11352 | 0.11693 | 0.07499 | 0.28715 |
| 0.4 | 0.1252 | | 0.172 | |
| 0.5 | 0.1407 | | 0.2438 | |
| 0.6 | 0.1623 | | 0.2855 | |
| 0.7 | 0.19278 | 0.17959 | 0.3130 | 0.34459 |
| 0.8 | 0.2378 | | 0.3442 | |
| 0.9 | 0.312 | | 0.4013 | |
| 0.99 | 0.4418 | | 0.5156 | |
| 0.995 | 0.4526 | | 0.5255 | |
| 0.999 | 0.4617 | | 0.5339 | |
| 1 | 0.46397 | 0.42033 | 0.53603 | 0.57967 |

Table III b

Albedo A* and diffuse transmittance T* of dielectric slabs (n=1.5) for isotropic surface illumination (Wu&Liou [11])

Double-layer albedo and transmittance
Isotropic source
$n_1$ = 1.333 $n_2$ = 1.5
$a_1 = a_2$ = 5

| $\varpi_1$ | $\varpi_2$ | A* K-int | A* Wu, Liou | T* K-int | T* Wu, Liou |
|---|---|---|---|---|---|
| 1 | 0.7 | 0.68974 | 0.68976 | 0.00374 | 0.00374 |
| 1 | 1 | 0.80927 | 0.80926 | 0.19073 | 0.19075 |
| 1 | 0.3 | 0.67968 | 0.67967 | 0.00036 | 0.00108 |
| 0.7 | 1 | 0.18501 | 0.18511 | 0.00437 | 0.00441 |
| 0.7 | 0.7 | 0.18489 | 0.18499 | 0.00019 | 0.00019 |
| 0.3 | 1 | 0.09515 | 0.09501 | 0.00034 | 0.00109 |
| 0.3 | 0.3 | 0.09515 | 0.09500 | 0.00001 | 0.00002 |

Table IVa

Double-layer albedo. K-integrals vs Wu&Liou [11] 8x8 matrix 12-dec calcs. Isotropic surface illumination $n_1$=1.333, $n_2$=1.5, $a_1$=$a_2$=5

| $\varpi_1$ | $\varpi_2$ | A* K-int | A* Wu, Liou | T* K-int | T* Wu, Liou |
|---|---|---|---|---|---|
| 1 | 0.7 | 0.65655 | 0.65897 | 0.00517 | 0.00441 |
| 0.7 | 1 | 0.1805 | 0.18046 | 0.00414 | 0.00374 |

Table IVb

Results for $n_1$=1.5, $n_2$=1.333, $a_1$=$a_2$=5

| $\varpi_1$ | $\varpi_2$ | A* asympt[(1)] | A* K-int | A* Wu, Liou | T* asympt[(1)] | T* K-int | T* Wu, Liou |
|---|---|---|---|---|---|---|---|
| 0.3 | 0.3 | 0.11216 | 0.11247 | 0.11245 | 0.00002 | 0.00001 | 0.00002 |
| 0.3 | 0.7 | 0.11216 | | 0.11245 | 0.00002 | | 0.00003 |
| 0.3 | 1 | 0.11218 | 0.11247 | 0.11246 | 0.00145 | 0.00038 | 0.0012 |
| 0.7 | 0.3 | 0.18377 | | 0.18033 | 0.00005 | | 0.00005 |
| 0.7 | 0.7 | 0.18377 | | 0.18034 | 0.00005 | | 0.00009 |
| 0.7 | 1 | 0.18389 | 0.18046 | 0.18045 | 0.00416 | 0.00468 | 0.0047 |
| 1 | 0.3 | 0.64019 | 0.63634 | 0.63632 | 0.00121 | 0.00054 | 0.00094 |
| 1 | 0.7 | 0.64329 | 0.6409 | 0.63924 | 0.00123 | 0.00164 | 0.00178 |
| 1 | 1 | 0.77335 | 0.77029 | 0.77008 | 0.22665 | 0.22971 | 0.22991 |

[(1)]Appendix II

Table IVc

Results for $n_1$=1.5, $n_2$=3, $a_1$=$a_2$=5

Half-space albedo vs incidence angle
n = 4/3 ϖ = 0.99

| θ* deg | A* diffusn | A* K-int | A* RTE | % error diffn | K-int |
|---|---|---|---|---|---|
| 0 | 0.6667 | 0.6526 | 0.6519 | 2.16 | -0.11 |
| 5 | 0.6667 | 0.6527 | 0.6521 | 2.14 | -0.09 |
| 10 | 0.6667 | 0.6533 | 0.6527 | 2.05 | -0.09 |
| 15 | 0.6668 | 0.6542 | 0.6536 | 1.93 | -0.09 |
| 20 | 0.6668 | 0.6554 | 0.6550 | 1.74 | -0.06 |
| 25 | 0.6669 | 0.6570 | 0.6567 | 1.51 | 0 |
| 30 | 0.6671 | 0.6590 | 0.6588 | 1.23 | 0 |
| 35 | 0.6675 | 0.6615 | 0.6613 | 0.91 | 0 |
| 40 | 0.6682 | 0.6644 | 0.6644 | 0.57 | 0 |
| 45 | 0.6693 | 0.6680 | 0.6681 | 0.19 | 0.01 |
| 50 | 0.6713 | 0.6725 | 0.6727 | -0.2 | 0.03 |
| 55 | 0.6746 | 0.6783 | 0.6786 | -0.5 | 0.04 |
| 60 | 0.6801 | 0.6863 | 0.6866 | -0.9 | 0.04 |
| 65 | 0.6895 | 0.6977 | 0.6981 | -1.2 | 0.06 |
| 70 | 0.7052 | 0.7150 | 0.7154 | -1.4 | 0.06 |
| 75 | 0.7321 | 0.7424 | 0.7428 | -1.4 | 0.05 |
| 80 | 0.7782 | 0.7877 | 0.7880 | -1.2 | 0.04 |
| 85 | 0.8583 | 0.8648 | 0.8650 | -0.8 | 0.02 |
| 90 | 1.0000 | 1.0000 | 1.0000 | 0 | 0 |

Table V

Albedo A* vs incident angle for planar illumination of a dielectric half-space n=1.333 333 ϖ=0.99. Diffusion, K-integral and RT results (Williams [15])

Slab albedo .
planar source/oblique
isotropic scattering
n = 1.338 $\varpi$ = 1

| $\mu^*$ | K-integral A* | Garcia A* | K-integral T* | Garcia T* |
|---|---|---|---|---|
| | d = 10 | | | |
| 0.2 | 0.863282 | 0.863546 | 0.136721 | 0.136454 |
| 0.5 | 0.810976 | 0.811236 | 0.189028 | 0.188764 |
| 1 | 0.78587 | 0.785552 | 0.214135 | 0.214448 |
| | d = 1 | | | |
| 0.2 | 0.43161 | 0.489159 | 0.34496 | 0.510842 |
| 0.5 | 0.24274 | 0.299713 | 0.53386 | 0.700287 |
| 1 | 0.20649 | 0.232583 | 0.79338 | 0.767417 |
| | d = 0.1 | | | |
| 0.2 | 0.30358 | 0.318291 | 0.473045 | 0.681709 |
| 0.5 | 0.044611 | 0.082988 | 0.732077 | 0.917012 |
| 1 | -0.087944 | 0.03893 | 1.087944 | 0.96107 |

Table VI

Albedo A* and diffuse transmittance T* for oblique planar illumination of dielectric layers on a non-scattering half-space (n=1.338) (Garcia [16])

n = 1.5  d = 10
isotropic incident light, isotropic scattering

| ϖ | A* diffusion | A* asympt. | A* K-integral | A* Wu, Liou | T* diffusion | T* asympt. | T* K-integral | T* Wu, Liou |
|---|---|---|---|---|---|---|---|---|
| 0 | 0.1558 | 0.09178 | 0.09178 | 0.09178 | | 0.00004 | | |
| 0.2 | 0.1668 | 0.10379 | 0.10428 | | | 0.00004 | 1.00E-06 | |
| 0.3 | 0.1741 | 0.11215 | 0.11247 | 0.11245 | | 0.000037 | 9.5E-006 | 0.00002 |
| 0.4 | 0.1831 | 0.12289 | 0.12271 | | | 0.000039 | 0.000024 | |
| 0.5 | 0.1948 | 0.13699 | 0.13603 | | | 0.000049 | 0.000044 | |
| 0.6 | 0.2105 | 0.15616 | 0.15413 | | | 0.000077 | 0.000081 | |
| 0.7 | 0.2332 | 0.18378 | 0.18036 | 0.18034 | 0.0004 | 0.000163 | 0.00018 | 0.00018 |
| 0.8 | 0.2699 | 0.22788 | 0.22268 | | 0.00063 | 0.00052 | 0.00057 | |
| 0.9 | 0.3442 | 0.31507 | 0.30803 | | 0.00376 | 0.0033 | 0.003545 | |
| 0.95 | 0.4298 | 0.41212 | 0.40493 | | 0.0157 | 0.0141 | 0.01479 | |
| 0.96 | 0.4588 | 0.44426 | 0.4373 | | 0.0226 | 0.0205 | 0.02144 | |
| 0.97 | 0.4964 | 0.48555 | 0.47902 | | 0.0342 | 0.0312 | 0.03256 | |
| 0.98 | 0.5487 | 0.54221 | 0.53641 | | 0.0551 | 0.051 | 0.05283 | |
| 0.99 | 0.6303 | 0.62924 | 0.62461 | | 0.09795 | 0.09221 | 0.09487 | |
| 0.995 | 0.6939 | 0.69634 | 0.69244 | | 0.1386 | 0.13195 | 0.13507 | |
| 0.999 | 0.7673 | 0.77333 | 0.76994 | | 0.1914 | 0.18431 | 0.18766 | |
| 1 | 0.7907 | 0.79781 | 0.79447 | 0.79445 | 0.2093 | 0.20219 | 0.20553 | 0.20555 |

Table VII

Comparison of diffusion, asymptotic (Appendix B), K-integral and RT results (Wu, Liou [11]) for albedo A* and diffuse transmittance T* vs scattering albedo ϖ for isotropic surface illumination of a dielectric slab (n=1.5, d=10)

Slab albedo vs boundary function
Isotropic source
L=10 n=1.5

| | $\varpi$ | asymp | $\Psi$ const | $\Psi$ log fn | $\beta$ | $1-\beta\mu$ | $\alpha$ | $1-\alpha\mu^2$ | Wu, Liou |
|---|---|---|---|---|---|---|---|---|---|
| A* | 1 | 0.79781 | 0.79403 | 0.79447 | 0.5 | 0.79445 | 0.43 | 0.79445 | 0.79445 |
| | 0.7 | 0.18378 | 0.17944 | 0.18036 | 0.52 | 0.18034 | 0.453 | 0.18034 | 0.18034 |
| | 0.3 | 0.11215 | 0.11223 | 0.11247 | 0.3 | 0.11245 | 0.17 | 0.11245 | 0.11245 |
| T* | 1 | 0.20219 | 0.20597 | 0.20553 | 0.5 | 0.20555 | 0.43 | 0.20555 | 0.20555 |
| | 0.7 | 0.00016 | 0.00018 | 0.00018 | 0.52 | 0.00018 | 0.453 | 0.00018 | 0.00018 |
| | 0.3 | 0.00004 | 0.00001 | 0.00001 | 0.3 | 0.00001 | 0.17 | 0.00001 | 0.00002 |
| L=1 n=1.5 | | | | | | | | | |
| A* | 1 | 0.48606 | 0.46093 | 0.46397 | 1.5302 | 0.42033 | 2.1521 | 0.42033 | 0.42033 |
| | 0.7 | 0.20969 | 0.1917 | 0.19278 | 1.425 | 0.17959 | 1.868 | 0.17959 | 0.17959 |
| | 0.3 | 0.1243 | 0.11327 | 0.11352 | 0.941 | 0.11693 | 0.8585 | 0.11693 | 0.11693 |
| T* | 1 | 0.51394 | 0.53907 | 0.53603 | 1.5302 | 0.57967 | 2.1521 | 0.57967 | 0.57967 |
| | 0.7 | 0.28442 | 0.31385 | 0.31300 | 1.29939 | 0.34459 | 1.5912 | 0.34459 | 0.34459 |
| | 0.3 | 0.2909 | | | | | | | 0.28715 |

Table VIII

K-integral results for slab albedo A* and diffuse transmittance T* vs choice of boundary transient cf RT data (Wu&Liou [11]) for isotropic surface illumination (n=1.5, d=10, 1)

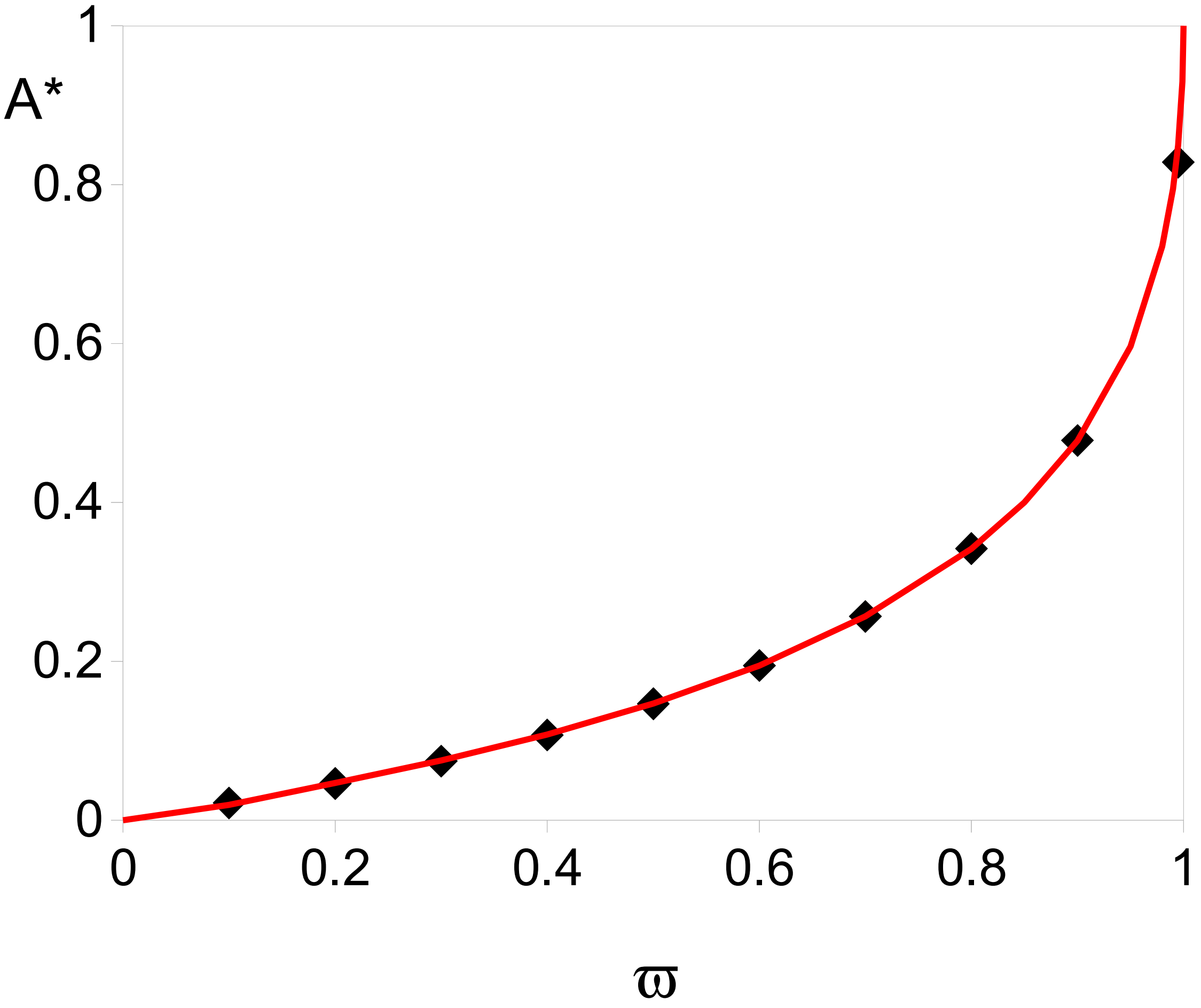

Fig 1

Half-space albedo A* vs particle scattering albedo ω (isotropic scattering, isotropic surface source, n = 1)  RT data (◆) [17] vs K-integral values (red curve)

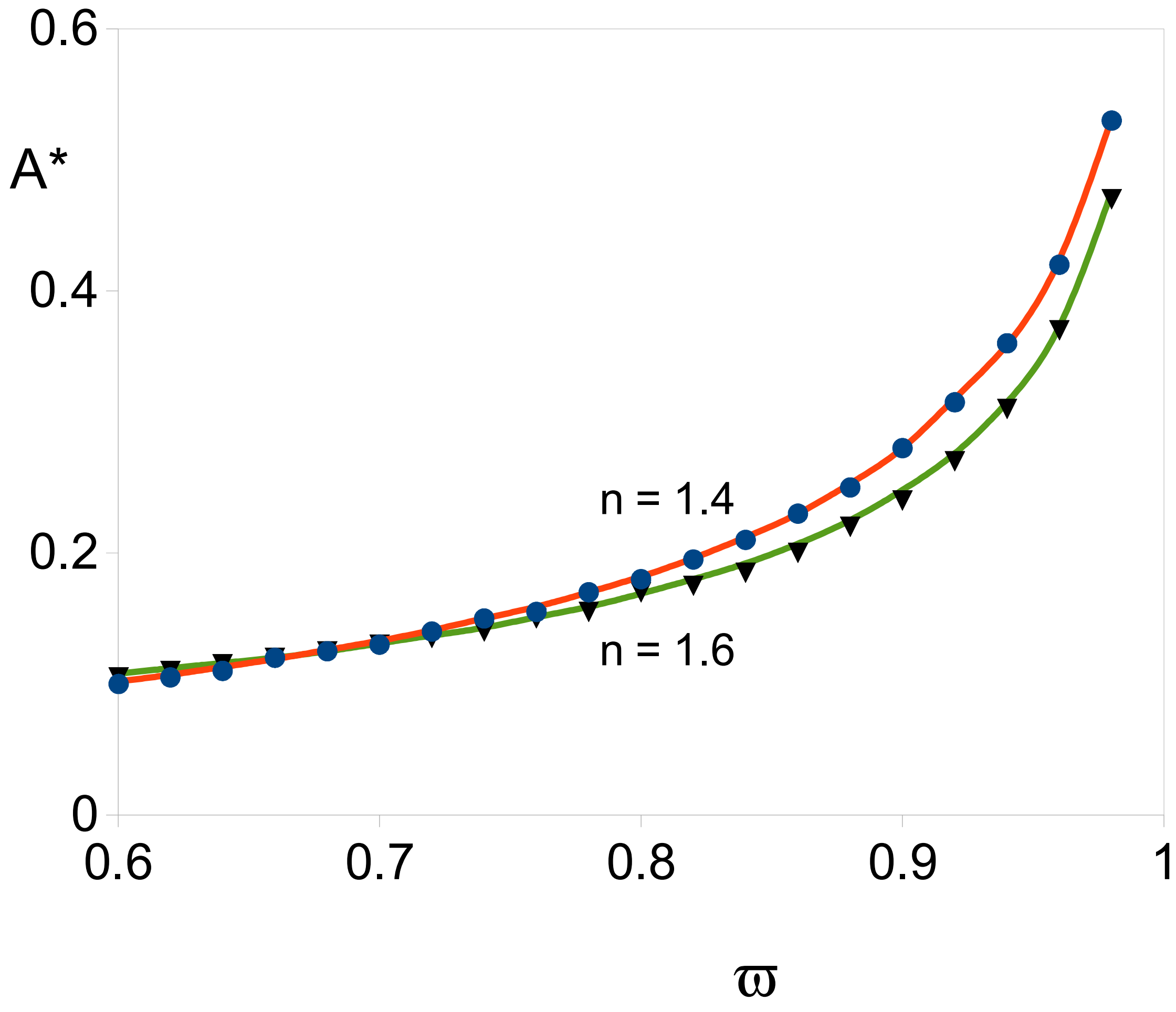

Fig 2

Half-space albedo A* vs scattering albedo $\varpi$ . Planar illumination at normal incidence (n = 1.4, 1.6). K-integral (smooth curves), RT data (●,▼) [18]

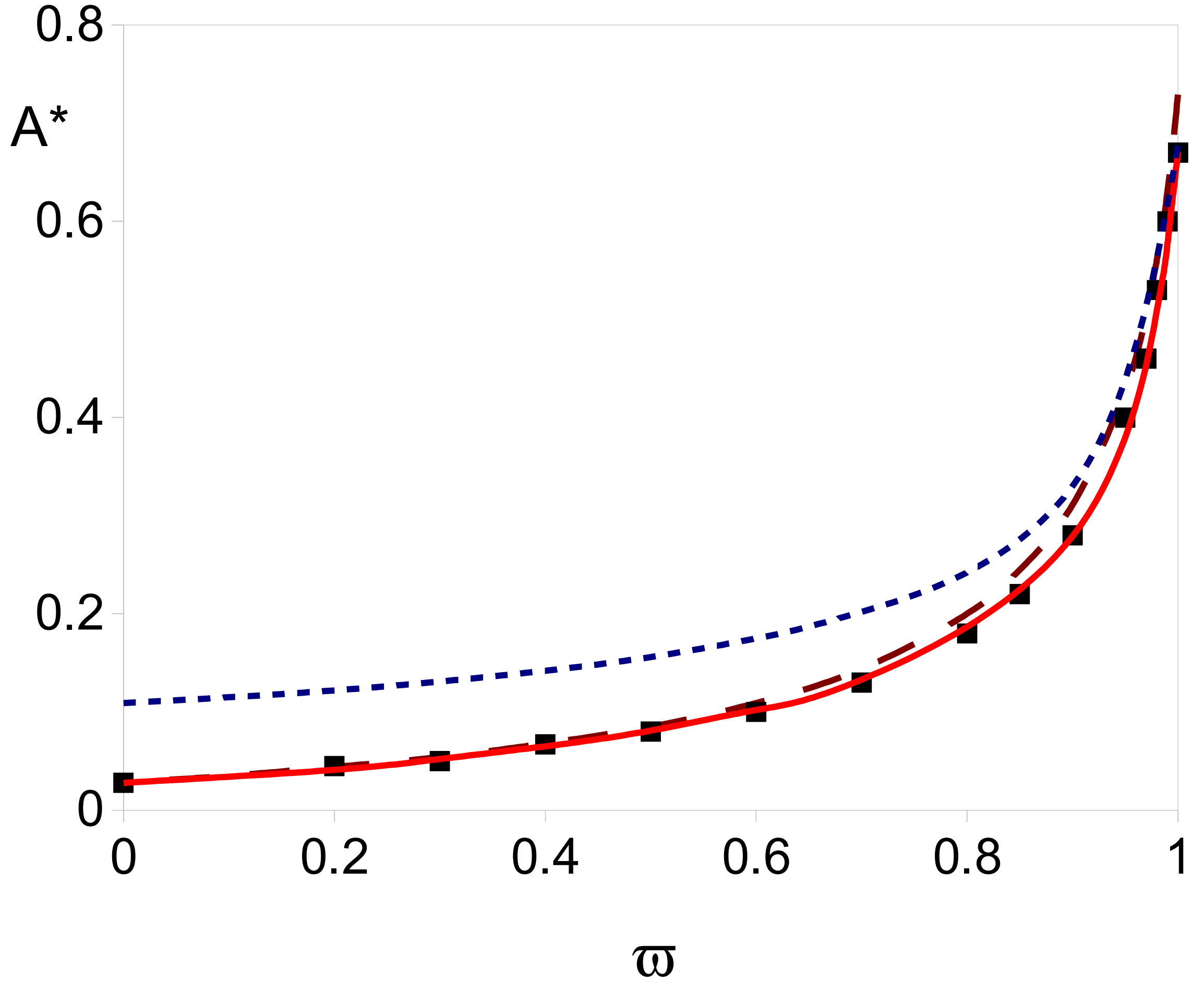

Fig 3

Slab albedo A* vs scattering albedo $\varpi$ . Diffusion (dotted curve), Dombrovsky 2-flux (dashed curve), K-integral (smooth curve), RT data (■) [18]. Planar illumination, normal incidence, isotropic scattering, slab width d = 5, n = 1.4

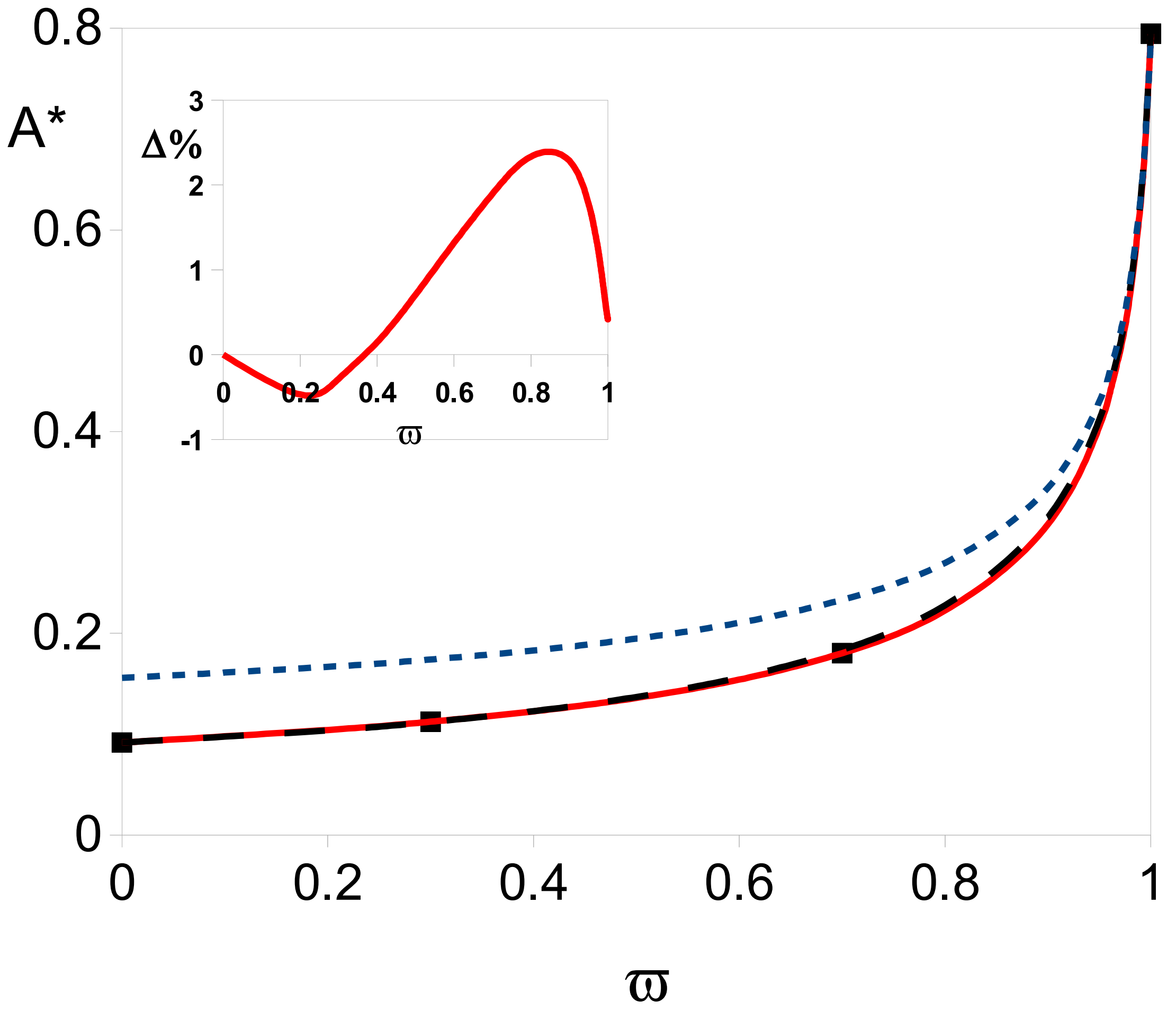

Fig 4

Slab albedo A* vs scattering albedo $\varpi$ . Diffusion (dotted curve), asymptotic (dashed curve), K-integral (smooth curve) vs RT data (■) [11] Isotropic illumination: d = 10, n = 1.5 (Table IIIa). Inset – percentage error of asymptotic vs K-integral results

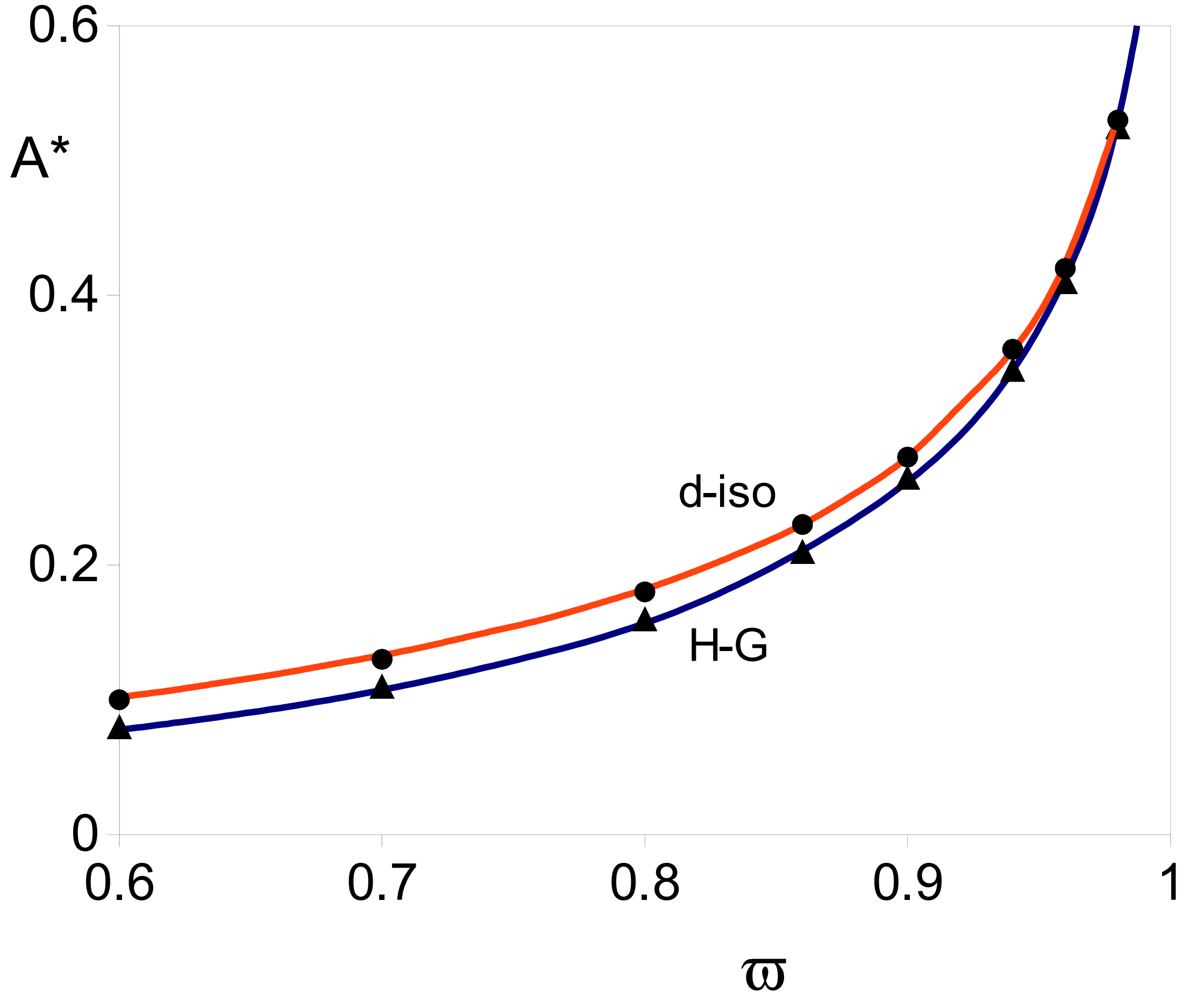

Fig 5

Half-space albedo A* vs scattering albedo $\varpi$, $\delta$-isotropic vs Henyey-Greenstein scattering.

Plane wave illumination, K-integral (smooth curves) vs RT (●,▲) [18]

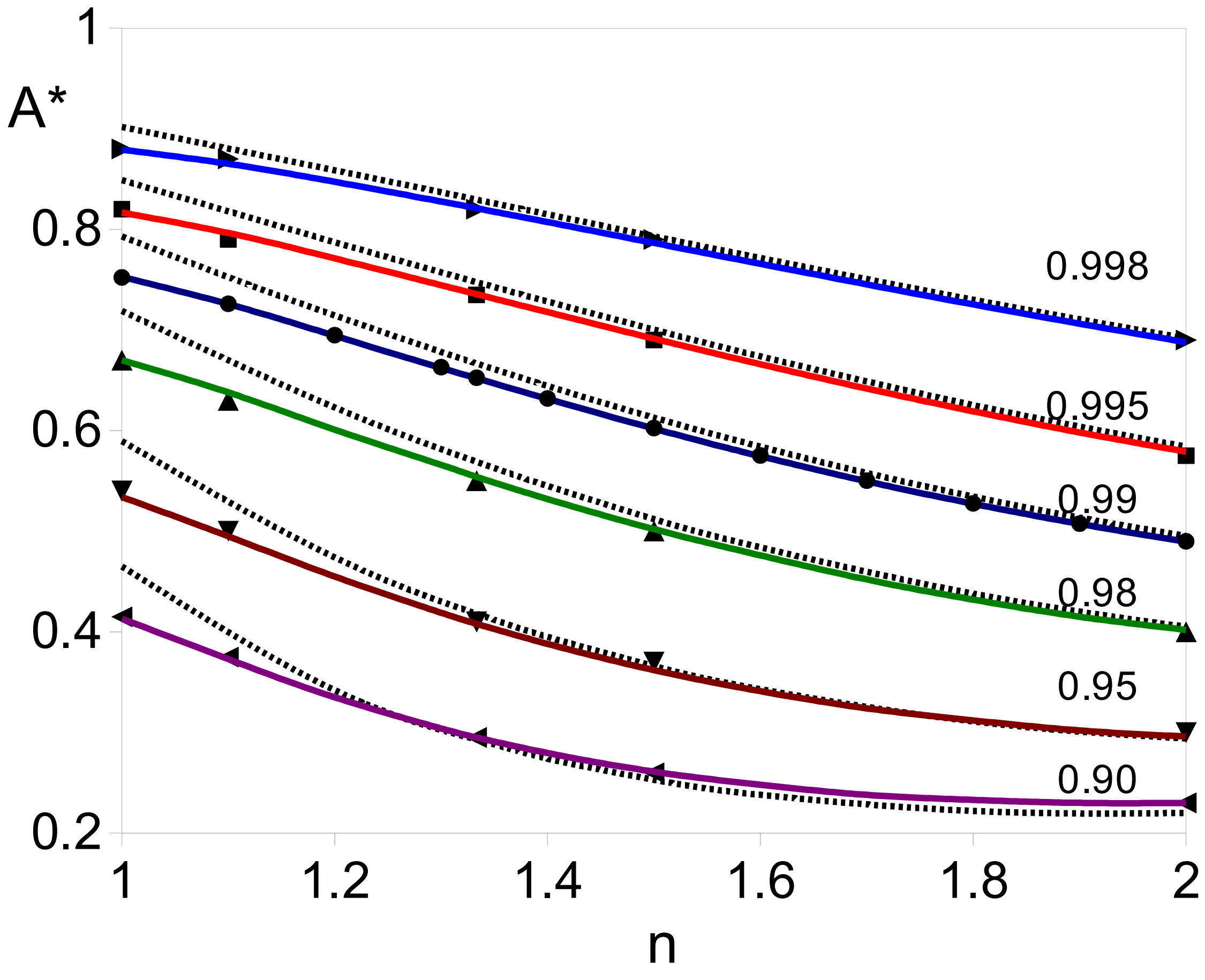

<u>Fig 6</u>

Half-space albedo A* vs refractive index n for a range of scattering albedoes. K-integral results (smooth curves) vs diffusion (dashed curves), RT values (filled symbols) [15]

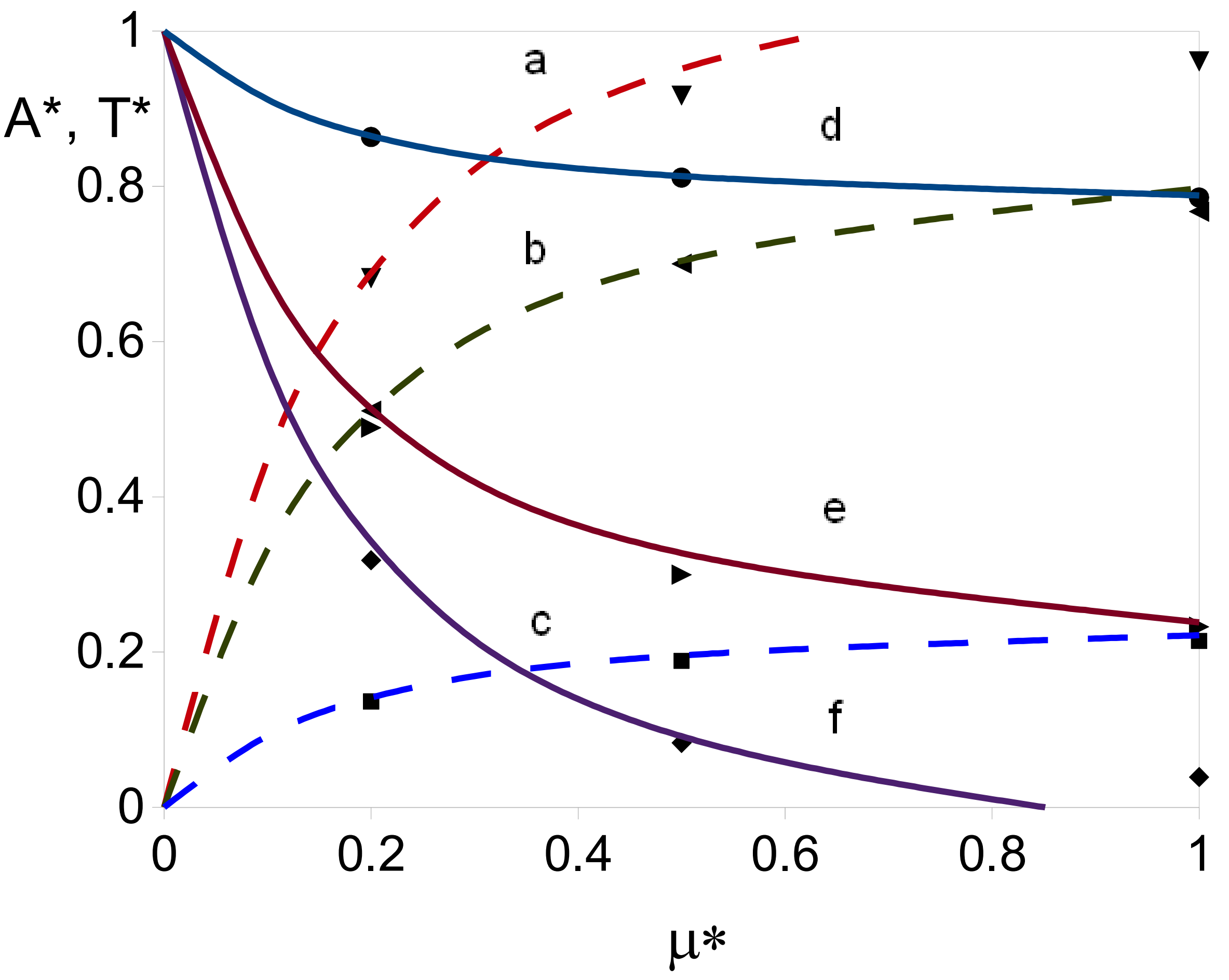

Fig 7

Slab albedo A* and transmittance T* vs cosine μ* for oblique planar illiumination. n = 1.338, d = 10, 1, 0.1 (isotropic scattering). K-integral results: A* - smooth curves, T* - dashed curves, RT data - filled symbols [16] d = 0.1 (a, f), 1 (b, e), 10 (c, f)

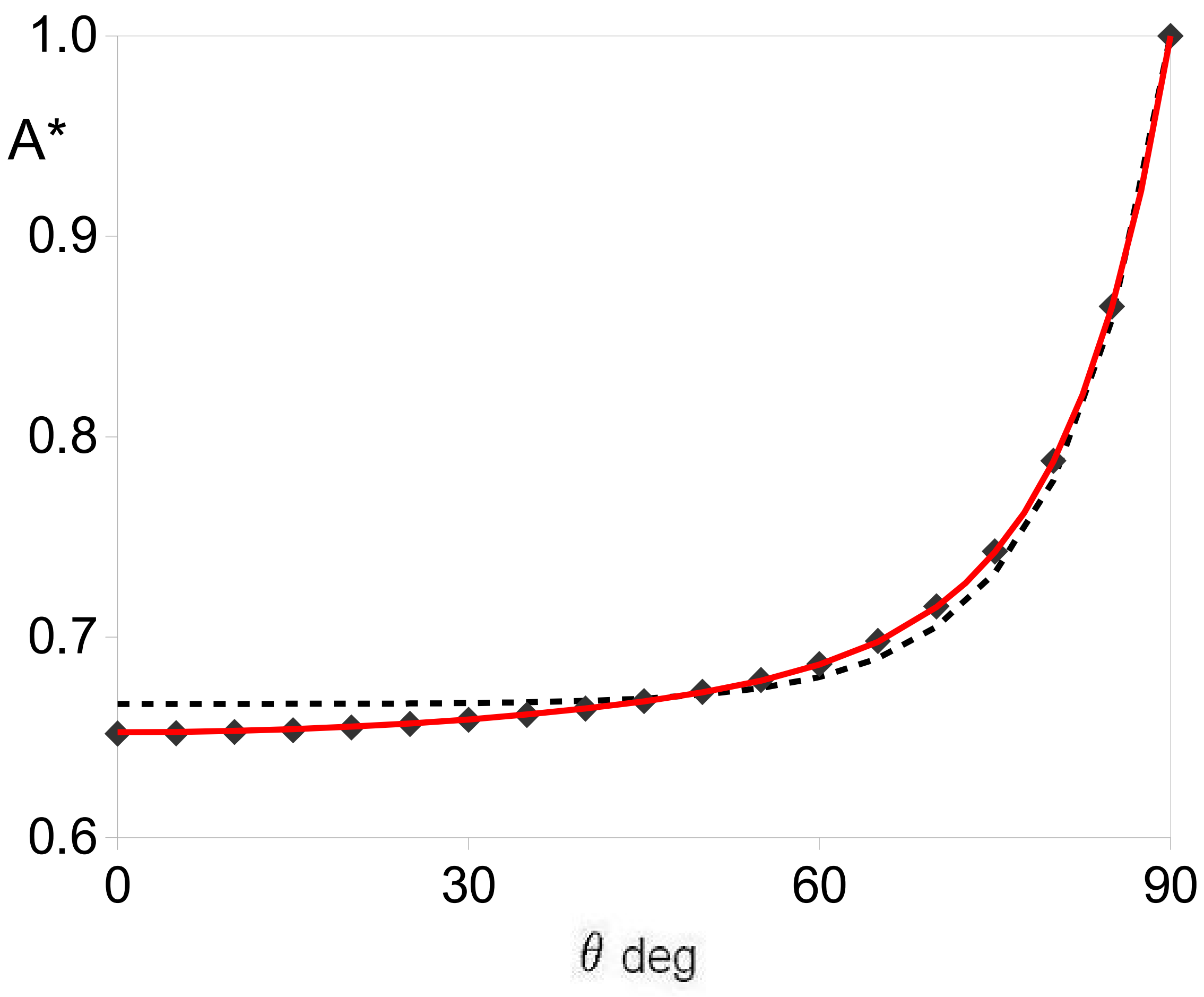

Fig 8

Albedo A* vs incident angle θ for plane-wave illumination of a half-space (ω = 0.99, n = 1.333333). K-integral results (smooth curve), diffusion data (dashed curve), RT data (◆) [15]

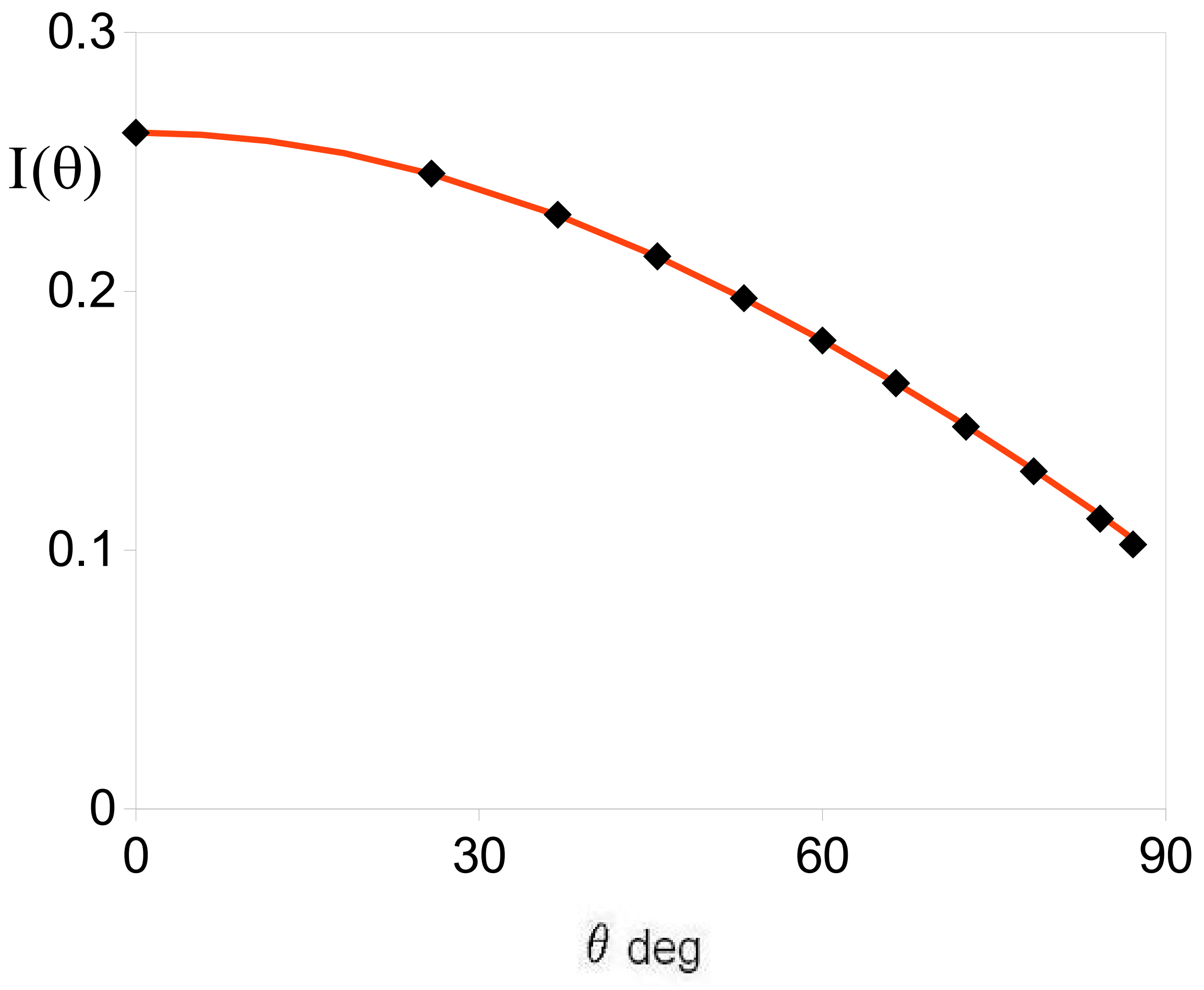

Fig 9

Forward angular intensity distribution I(d,θ) at the exit face for slab width d =5. Isotropic incident light, scattering albedo ω = 1, n = 1. Transport data (◆) [12, 13] vs K-integral results (red curve)

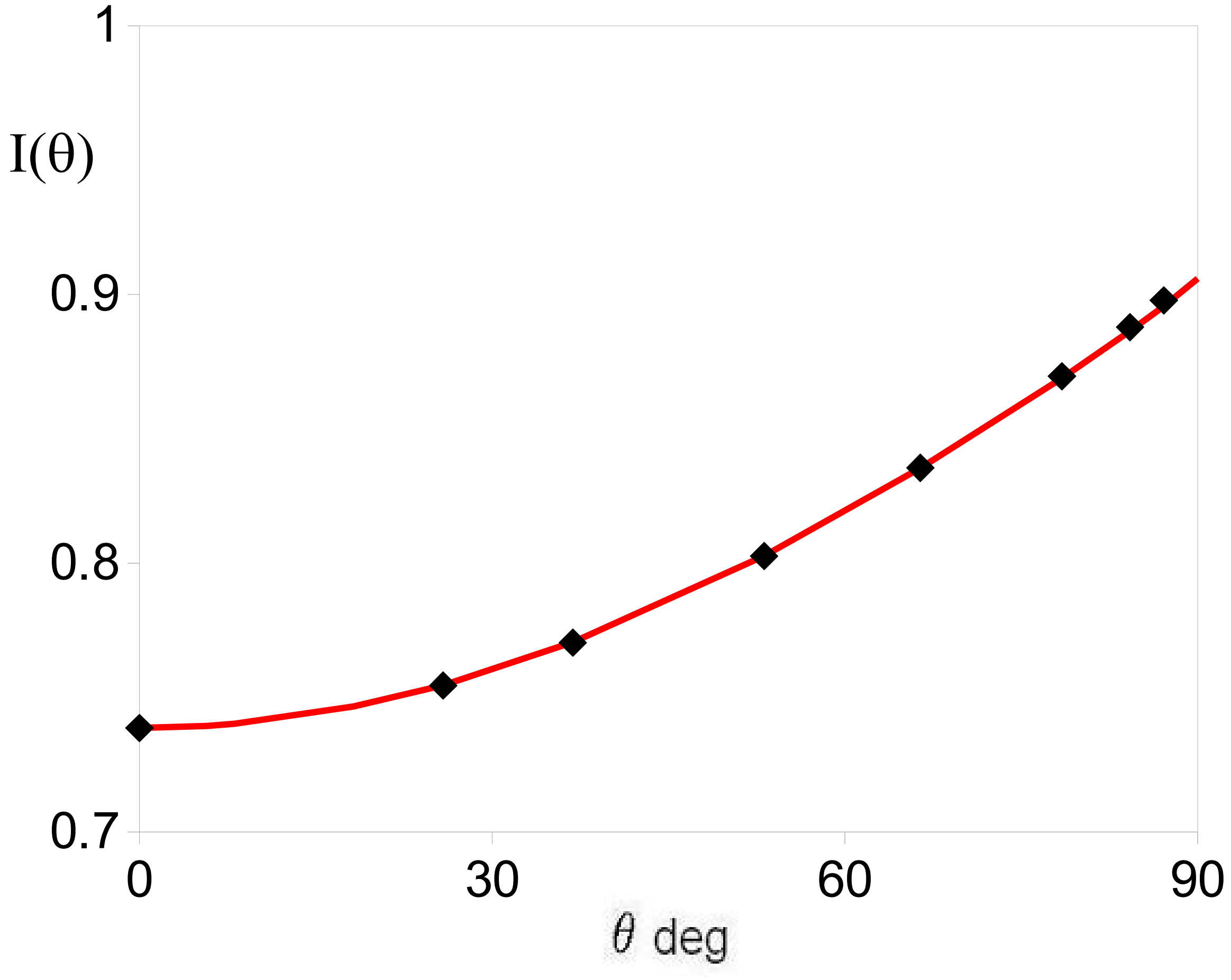

Fig 10

Backward angular intensity I(0,-θ) at the illuminated surface of a slab of thickness d = 5. Isotropic incident light, scattering albedo ω = 1, n = 1. Transport data (◆) [12, 13] vs K-integral results (red curve)

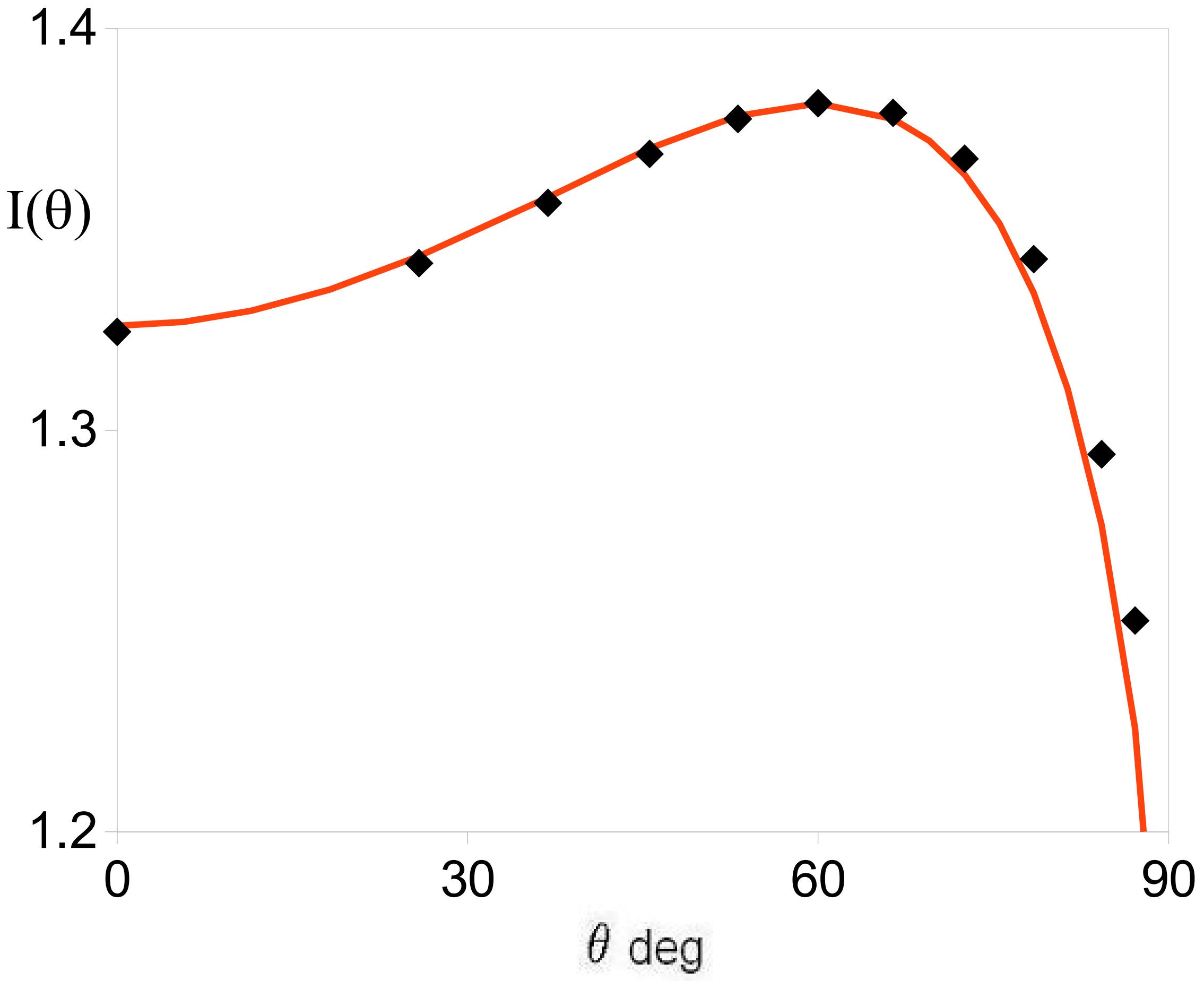

Fig 11

Backward angular intensity I(0,-θ) emitted at the illuminated surface of a slab of width d = 5, plane wave source incident at $\mu^* = 0.9$, scattering albedo $\omega = 1$, n = 1. Transport data (◆) [12, 13] vs K-integral results (red curve)

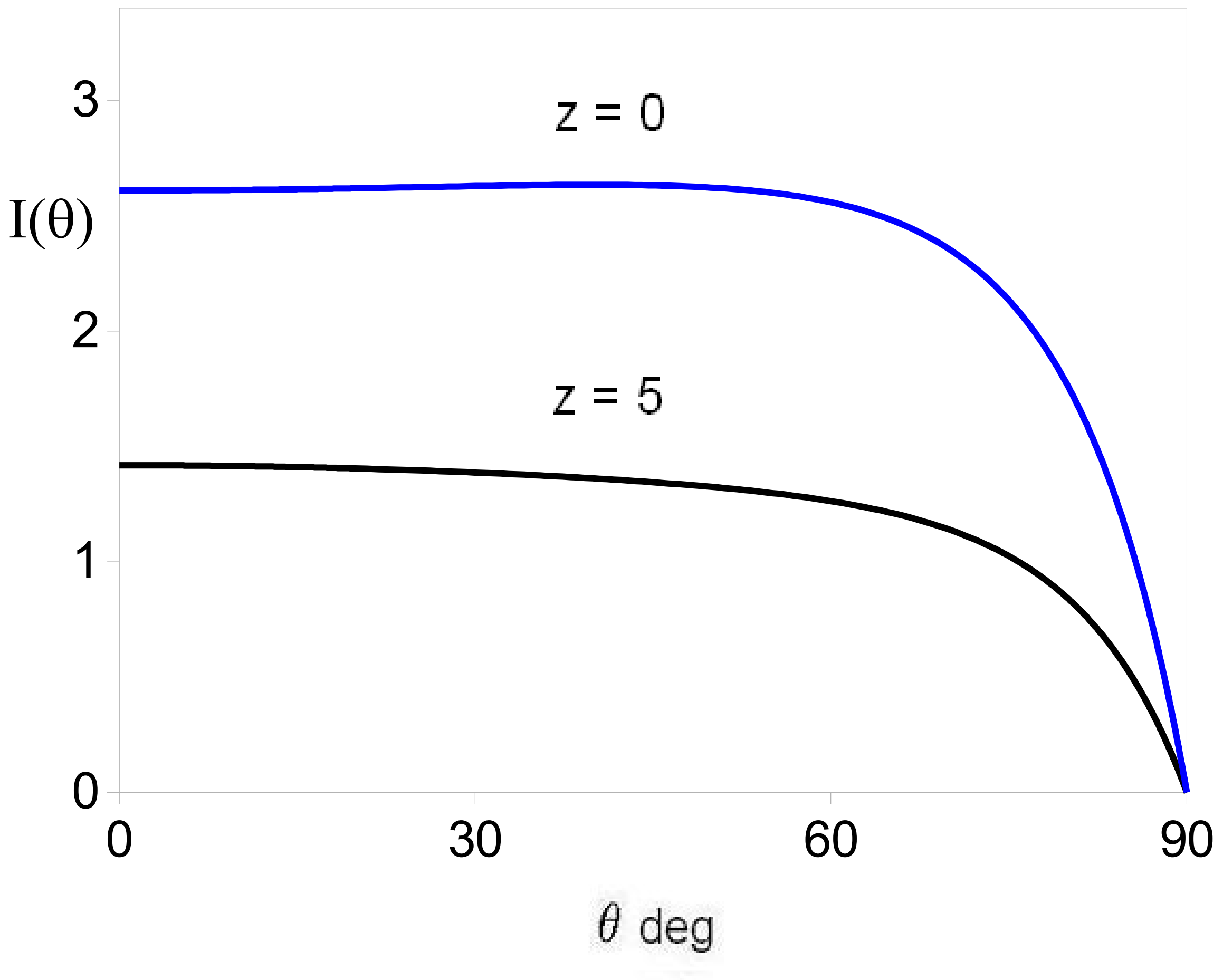

Fig 12

Slab angular emission intensity I(θ): planar source, normal incidence (n=1.4, d=5,ϖ =1)

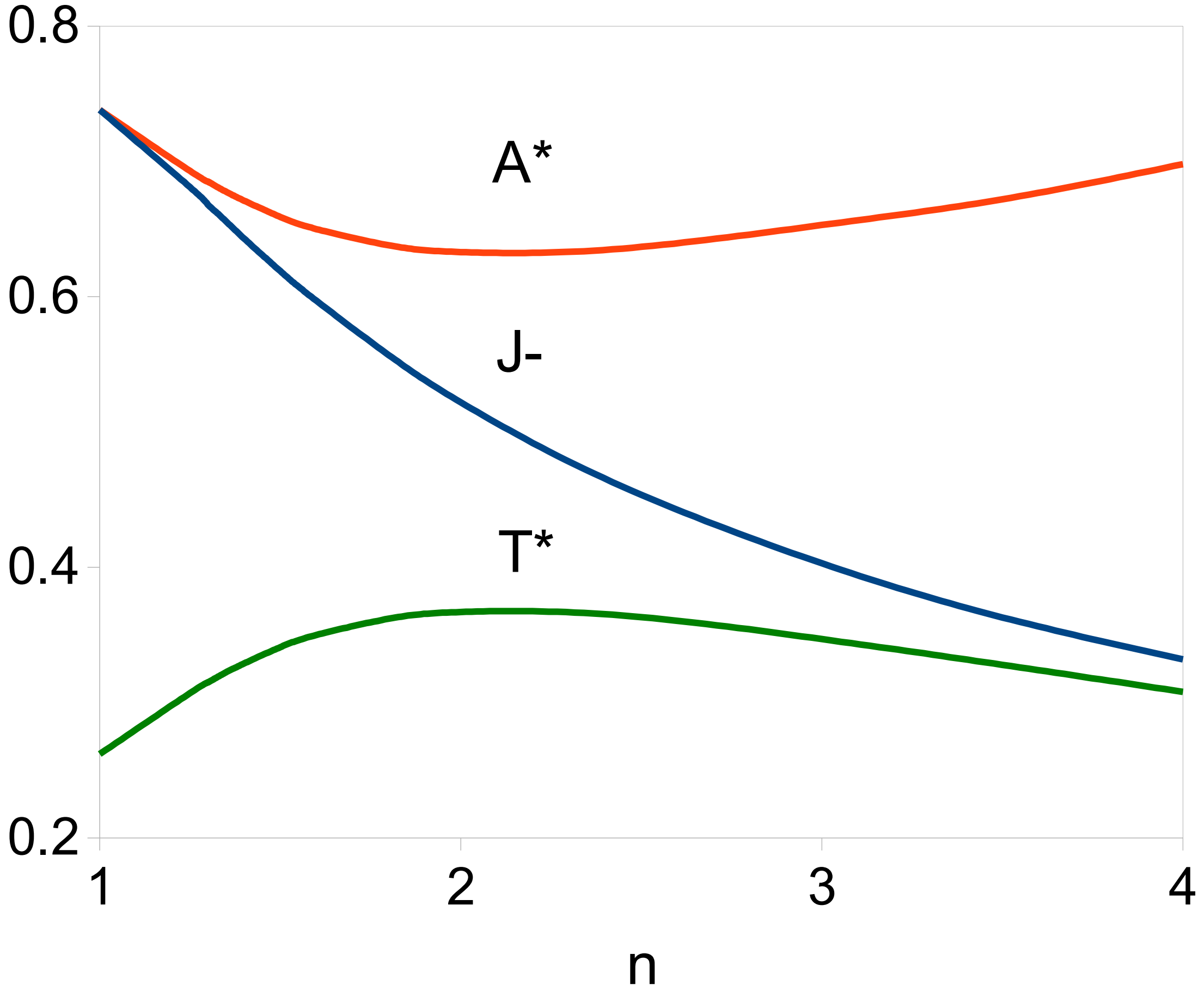

Fig 13

Slab albedo A*, diffuse transmittance T*, exit flux J- at z=0 vs refractive index n

Planar source, normal incidence. K-integral results for slab width d=5 (ϖ =1)

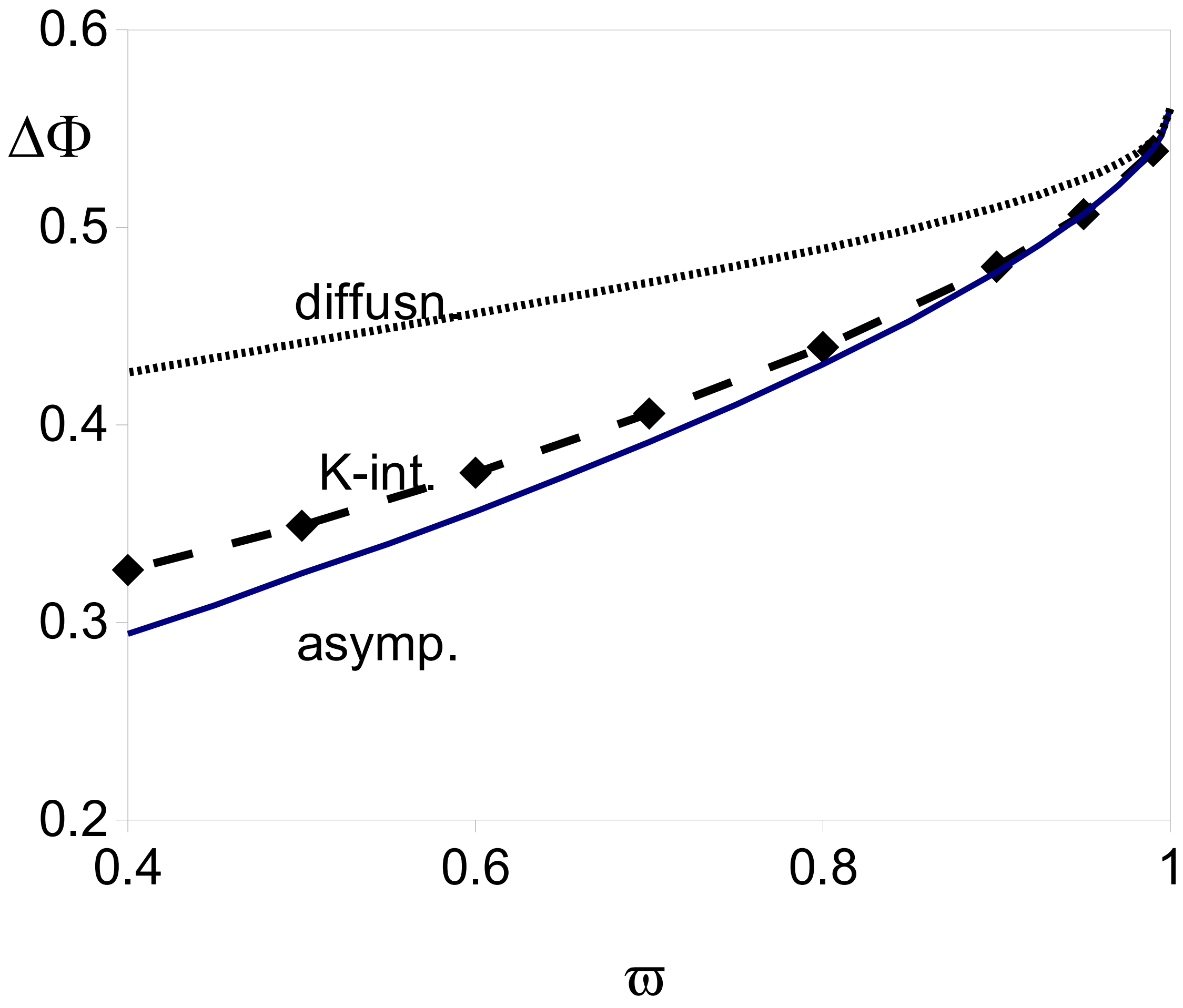

Fig 14

Comparison of diffusion, asymptotic transport and K-integral results for flux density jump at the refractive index discontuity at an interface between two scattering media vs scattering albedo $\varpi$ (refractive index ratio n = 4/3, isotropic scattering) [4]